\documentclass{aa}  
\usepackage{rotating}
\usepackage{hyperref}
\usepackage{epstopdf}
\usepackage{graphicx}
\usepackage{overpic}

\usepackage{multirow}
\usepackage{txfonts}
\begin{document}

   \title{Modelling a multi-spacecraft coronal mass ejection encounter with EUHFORIA}

   \subtitle{}

   \author{E. Asvestari
          \inst{1}
          \and
          J. Pomoell\inst{1}
          \and
          E. Kilpua\inst{1}
          \and
          S. Good\inst{1}
          \and
          T. Chatzistergos\inst{2}
          \and
          M. Temmer\inst{3}
          \and
          E. Palmerio\inst{4,5}
          \and
          S. Poedts\inst{6,7}
          \and
          J. Magdalenic\inst{6,8}
          }

   \institute{University of Helsinki, 00100 Helsinki, Finland\\
              \email{eleanna.asvestari@helsinki.fi}
        \and
            Max Planck Institute for Solar System Research, Justus-von-Liebig-Weg 3, 37077 G{\"o}ttingen, Germany
         \and
            Institute of Physics, University of Graz, 8010 Graz, Austria
            \and
            Space Sciences Laboratory, University of California--Berkeley, Berkeley, CA 94720, USA
            \and
            CPAESS, University Corporation for Atmospheric Research, Boulder, CO 80301, USA
            \and
            Centre for mathematical Plasma Astrophysics, KU Leuven, 3001 Leuven, Belgium
            \and
            Institute of Physics, University of Maria Curie-Sk{\l}odowska, ul.\ Radziszewskiego 10, PL-20-031 Lublin, Poland
            \and
            Solar-Terrestrial Centre of Excellence – SIDC, Royal Observatory of Belgium, 1180 Brussels, Belgium
             }

   \date{Received ; accepted }

 
  \abstract
   {Coronal mass ejections (CMEs) are a manifestation of the Sun’s eruptive nature. They can have a great impact on Earth, but also on human activity in space and on the ground. Therefore, modelling their evolution as they propagate through interplanetary space is essential.
   }
   {EUropean Heliospheric FORecasting Information Asset (EUHFORIA) is a data-driven, physics-based model, tracing the evolution of CMEs through background solar wind conditions. It employs a spheromak flux rope, which provides it with the advantage of reconstructing the internal magnetic field configuration of CMEs. This is something that is not included in the simpler cone CME model used so far for space weather forecasting. This work aims at assessing the spheromak CME model included in EUHFORIA. 
   }
   {We employed the spheromak CME model to reconstruct a well observed CME and compare model output to in situ observations. We focus on an eruption from 6 January 2013 that was encountered by two radially aligned spacecraft, Venus Express and STEREO-A. We first analysed the observed properties of the source of this CME eruption and we extracted the CME properties as it lifted off from the Sun. Using this information, we set up EUHFORIA runs to model the event.
   }
   {The model predicts arrival times from half to a full day ahead of the in situ observed ones, but within errors established from similar studies. In the modelling domain, the CME appears to be propagating primarily southward, which is in accordance with white-light images of the CME eruption close to the Sun.
   }
   {In order to get the observed magnetic field topology, we aimed at selecting a spheromak rotation angle for which the axis of symmetry of the spheromak is perpendicular to the direction of the polarity inversion line (PIL). The modelled magnetic field profiles, their amplitude, arrival times, and sheath region length are all affected by the choice of radius of the modelled spheromak.
   }

   \keywords{Sun: coronal mass ejections (CMEs) – Sun: heliosphere – Sun: magnetic fields – solar-terrestrial relations – solar wind – magnetohydrodynamics (MHD)
               }

   \maketitle
%

\section{Introduction}\label{intro}

Coronal mass ejections (CMEs) are enormous plasma clouds ejected from the solar corona with velocities that can reach up to 3000 km/s and a mass that can be up to a few $10^{16}$ g  \citep[e.g.][]{webb_coronal_2012}. During their journey through the heliosphere, they can be a potential hazard for human health and activity in space and on the ground \citep[e.g.][and references therein]{lanzerotti_space_2001, lanzerotti_space_2001-1, daglis_effects_2004,hapgood_towards_2011,cannon_extreme_2013, green_coronal_2015, schrijver_understanding_2015, eastwood_economic_2017}. As an interplanetary CME (ICME) expands in space, it carries helical magnetic field lines with it, in the form of a flux rope, which are frozen in the CME plasma \citep{webb_coronal_2012}. At this point, it is worth noting that not all ICMEs show evidence of an embedded flux rope \citep{vourlidas_how_2013}, with this being more common during periods of solar maximum, according to \citet{cane_interplanetary_2003}. The structure of the flux rope, namely the configuration of its magnetic field components and the magnitude of the field, are important elements when assessing the impact of a CME \citep[e.g.][]{kilpua_geoeffective_2017}. Historically, (semi)-empirical and physics-based CME forecasting models have primarily focussed on predicting the arrival time of CMEs at Earth, as well as details of the encounter, for example whether it was a nose or flank encounter \citep{ mays_ensemble_2015, riley_forecasting_2018}. Their accuracy on determining the arrival time on Earth is of the order of 10 hours ahead or after the actual CME arrival with some extreme cases of 3 hours up to even several days \citep[see][and references therein]{zhao_current_2014, mays_ensemble_2015, paouris_effective_2017, riley_forecasting_2018, verbeke_benchmarking_2019}. These models so far make no prediction for the magnetic field configurations of CMEs arriving at Earth. Recent developments in existing models include magnetized CMEs, thus presenting a significant potential to improve space weather forecasts  \citep{manchester_flux_2014, isavnin_fried_2016, shiota_magnetohydrodynamic_2016, jin_data-constrained_2017, jin_chromosphere_2017, kay_effects_2018, verbeke_evolution_2019, scolini_observation-based_2019, scolini_cmecme_2020, kay_fido-sit_2020}.

An aspect necessary to consider in forecasting models is that CMEs actively interact with the ambient solar wind and structures embedded within it, in particular with slow--fast stream interaction regions and other CMEs \citep[see][]{manchester_physical_2017}. This can significantly impact the CME evolution and propagation \citep{macqueen_propagation_1986, isavnin_three-dimensional_2014}. From their onset and throughout their journey, CMEs and their embedded flux ropes can undergo deformation, kink, rotation, deflection, and erosion through reconnection \citep{kay_global_2015, kay_heliocentric_2015, heinemann_cmehss_2019}. Regardless of whether these types of processes take place close to the Sun or further out in interplanetary space, they affect the spatial and magnetic field configuration of the CME, complicating forecasts of its arrival and geoeffectiveness \citep{mostl_strong_2015}. It is, therefore, important in determining the global success of forecasting models to use past CME events that have been observed in situ at different heliodistances.

The aim of the current paper is to assess the spheromak CME model included in the EUropean Heliospheric FORecasting Information Asset \citep[EUHFORIA;][]{pomoell_euhforia_2018, verbeke_evolution_2019} by comparing the model output to multi-point in situ observations. We focus on the model’s capability to predict the arrival time of the CME and the temporal profiles of its magnetic field magnitude and components, as well as on  estimating the evolution of the CME in interplanetary space. For this purpose, a CME that has been well observed by multiple spacecraft at varying heliospheric distances was chosen following the selection criteria described in Section \ref{Cand_Selec}. The CME eruption is estimated to have occurred at around 03:30UT on 6 January 2013. Clear flux rope signatures were observed in situ by \emph{Venus Express} and \emph{Solar TErrestrial RElations Observatory-A} (STEREO-A) spacecraft. The \emph{MErcury Surface, Space ENvironment, GEochemistry, and Ranging} (MESSENGER) spacecraft, located in the vicinity of the other two (26.8$^{\circ}$ apart from Venus Express in longitude), also registered some minor disturbance but without a clear flux rope signature. The remote-sensing observations of the eruption are discussed in Section \ref{remotesensing}. The graduated cylindrical shell (GCS) method was applied to obtain the location, geometry, and kinematic parameters of the CME, as described in Sections \ref{properties}, which were then used as input parameters for EUHFORIA, as detailed in Section \ref{model}. In the same section the output is compared to in situ signatures at the spacecraft and the results are discussed. From this analysis, it can be concluded that many aspects of the model output, in particular the propagation direction, are highly sensitive to the white-light images that will be selected for the GCS analysis. For the CME studied, the GCS reconstruction implied that the CME apex propagated clearly southward, and that only the CME flank would have intersected the solar equatorial (SE) plane. However, Venus Express and STEREO-A, which are both located near the SE plane, observed clear flux rope rotations consistent with an encounter closer to the apex. Potential causes of this discrepancy are discussed in detail in Section \ref{model}.


\section{Databases} \label{databases}

In this study, we employed remote-sensing observations of the Sun and in situ observations by the \emph{Solar and Heliospheric Observatory} \citep[SOHO; ][]{domingo_soho_1995}, the \emph{Solar Dynamics Observatory} \citep[SDO; ][]{pesnell_solar_2012}, the \emph{Solar Terrestrial Relations Observatory} \citep[STEREO; ][]{kaiser_stereo_2005, kaiser_stereo_2008},
and the \emph{Venus Express} \citep{titov_venus_2006, svedhem_venus_2007} missions. More precisely, for investigating the source of the CME, we considered extreme ultraviolet (EUV) filtergrams obtained by the Atmospheric Imaging Assembly \citep[AIA; ][]{lemen_atmospheric_2012} instrument on board SDO, and the Sun Earth Connection Coronal and Heliospheric Investigation \citep[SECCHI; ][]{howard_sun_2002, howard_sun_2008} Extreme UltraViolet Imager (EUVI) instrument on board STEREO-A \& B. In addition, we explored the magnetic field topology using the Helioseismic and Magnetic Imager \citep[HMI; ][]{scherrer_helioseismic_2012} on board SDO. We analysed the CME signatures in the corona using white-light coronagraph images taken by the Large Angle and Spectrometric COronagraph \citep[LASCO; ][]{brueckner_large_1995} C2 \& C3 telescopes on board SOHO and the SECCHI--COR1 \& SECCHI--COR2 coronagraphs on board the two STEREO spacecraft. The filament features were also examined using full-disc H$\alpha$ images of the Sun taken with the solar telescope at Kanzelh{\"o}he Solar Observatory.

To compare the model output to in situ signatures, we used plasma and magnetic field measurements in the solar wind made by STEREO-A and Venus Express, curated by the \emph{Heliospheric Cataloguing, Analysis and Techniques Service} \citep[HELCATS;][]{harrison_cmes_2018} project Work Package (WP) 4 DATACAT products. In search of our test candidates, we surveyed the HELCATS WP4 linked catalogue (\url{https://www.helcats-fp7.eu/catalogues/wp4_cat.html}), as well as the list of ICME signatures identified at radially aligned spacecraft created by \cite{good_self-similarity_2019} and the Earthbound CME list given in \cite{palmerio_coronal_2018}.

EUHFORIA employs magnetograms in order to provide the inner boundary conditions at 0.1 AU and to reconstruct the ambient solar wind conditions. The magnetogram used is provided by the Global Oscillation
Network Group (GONG). For the purpose of this study, we employed
the recently updated Air Force Data Assimilative Photospheric
Flux Transport \citep[ADAPT; ][]{arge_air_2010} magnetograms (\url{ftp://gong2.nso.edu/adapt/maps/gong/}).


\begin{figure}
   \centering
   \includegraphics[width=0.90\hsize]{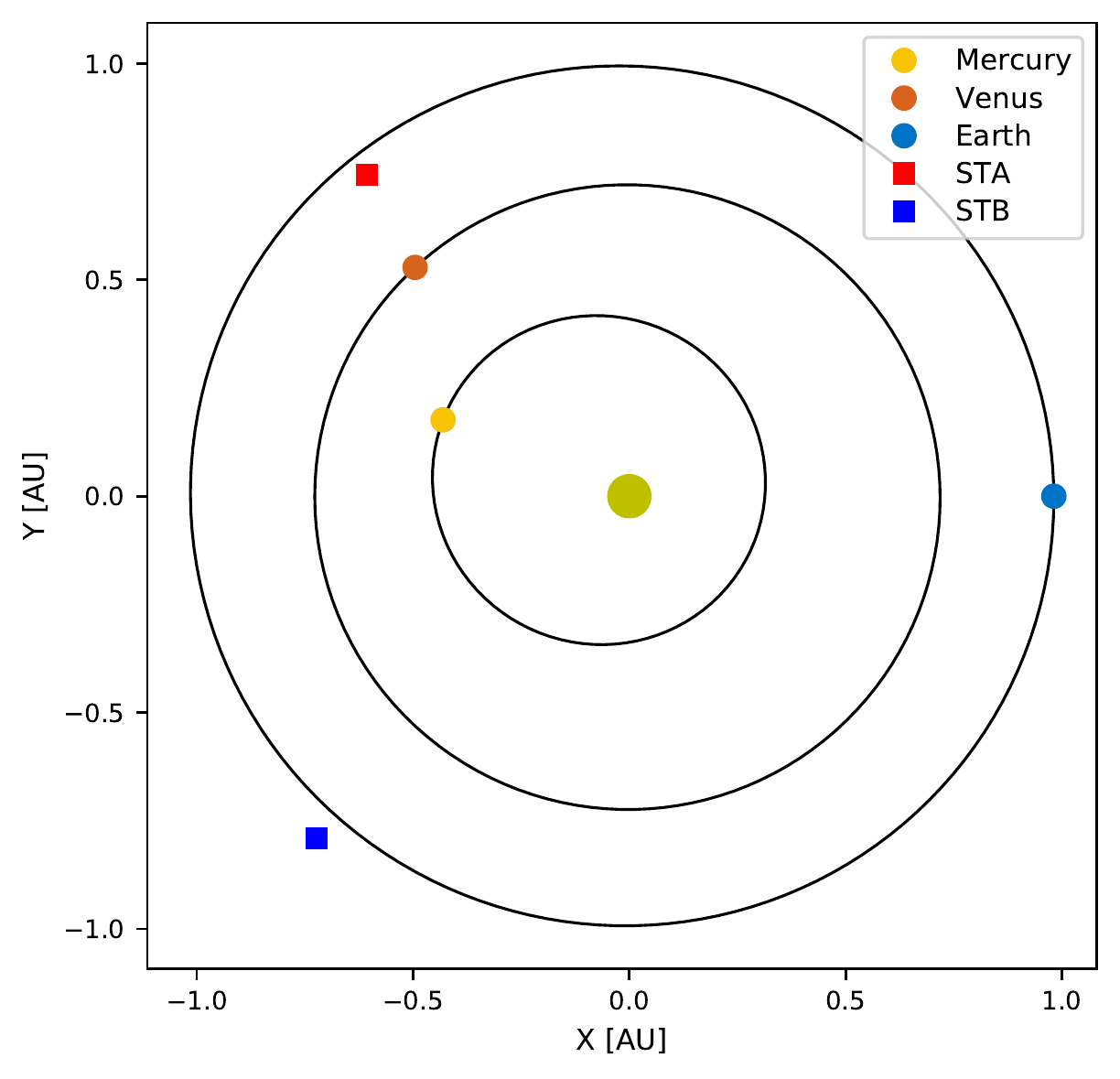}
      \caption{Spacecraft locations during the CME passage from Venus Express to STEREO-A. MESSENGER was also nearby and recorded some disturbance without registering a clear flux rope.
              }
         \label{orbit}
   \end{figure}


\section{CME candidate selection}\label{Cand_Selec}

For the purpose of this study, we aimed to select a well isolated and clear CME event that was observed at two locations in the inner heliosphere by well separated spacecraft. Therefore, CMEs that were immediately preceded by, followed by, or interacting with other CMEs had to be excluded. Also, events with complex or vague sources, and with complicated or ambiguous in situ signatures were also not considered. In addition, CME events for which the in situ observations contain data gaps that impede the identification of the event arrival time and duration, as well as the determination of its magnetic field configuration at the spacecraft, were excluded. These limitations resulted in a very small selection of possible CME events, out of which one was singled out as the best candidate, satisfying all criteria.

The selected CME eruption took place during the maximum phase of Solar Cycle 24 on 6 January 2013 around 03:30UT as indicated by EUV images from STEREO-B and SDO. More details on the EUV filtergrams is given in section \ref{remotesensing}. The eruption was treated as an isolated event, that is to say no other EUV eruptions were detected close by at the centre of the STEREO-A field of view, even though another CME signature was present in the remote sensing observations. Preconditioning of the interplanetary space and possible interaction cannot be fully excluded, however, as discussed in Appendix \ref{cone_EUHFORIA}, in the modelling domain the two ICMEs did not appear to interact. This encourages us to focus on modelling a single CME. Based on the  spacecraft positions (Figure \ref{orbit}), the ICME was expected to be encountered by STEREO-A and Venus Express. In situ measurements confirm that similar flux rope signatures were observed by both spacecraft \citep[see][for a detailed analysis]{good_self-similarity_2019, vrsnak_heliospheric_2019}. At MESSENGER, only a shock-like discontinuity without any significant magnetic field strength, |B|, enhancement was registered. Based on the spacecraft position and the tilt of the erupted filament, a near flank encounter of the ICME with MESSENGER was anticipated. At the time of the cloud passage, Venus Express and STEREO-A were radially aligned (Figure \ref{orbit}), making this event ideal for the study. The longitudinal and latitudinal separations between Venus Express and STEREO-A were 3.8$^{\circ}$and 2.4$^{\circ}$, respectively, and between Venus Express and MESSENGER the separations were 26.8$^{\circ}$ and 3$^{\circ}$, respectively.

\begin{figure}
    \centering
    \begin{minipage}{0.49\textwidth}
    \centering
        \includegraphics[width = 1\textwidth]{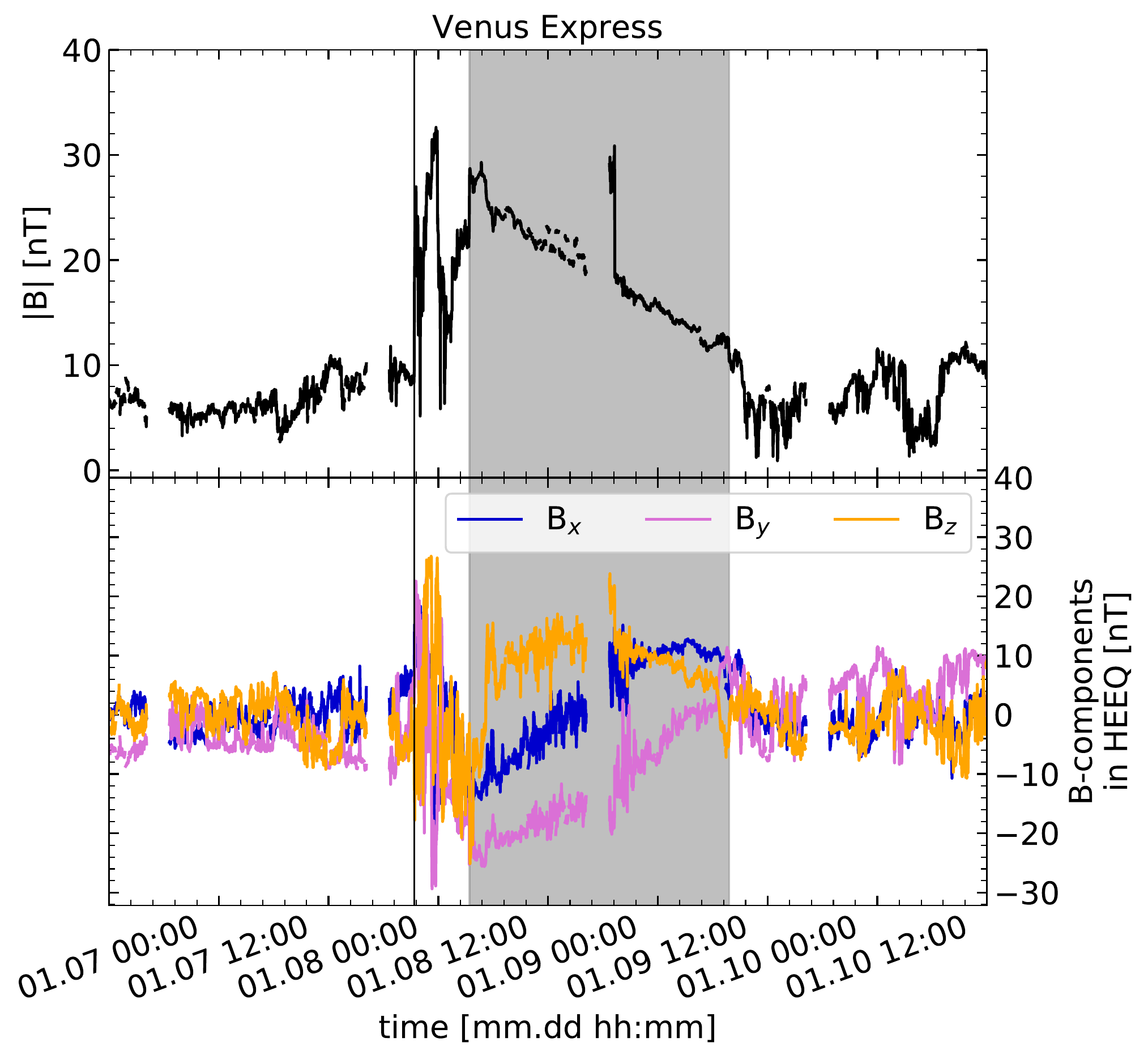}
        \includegraphics[width = 0.96\textwidth]{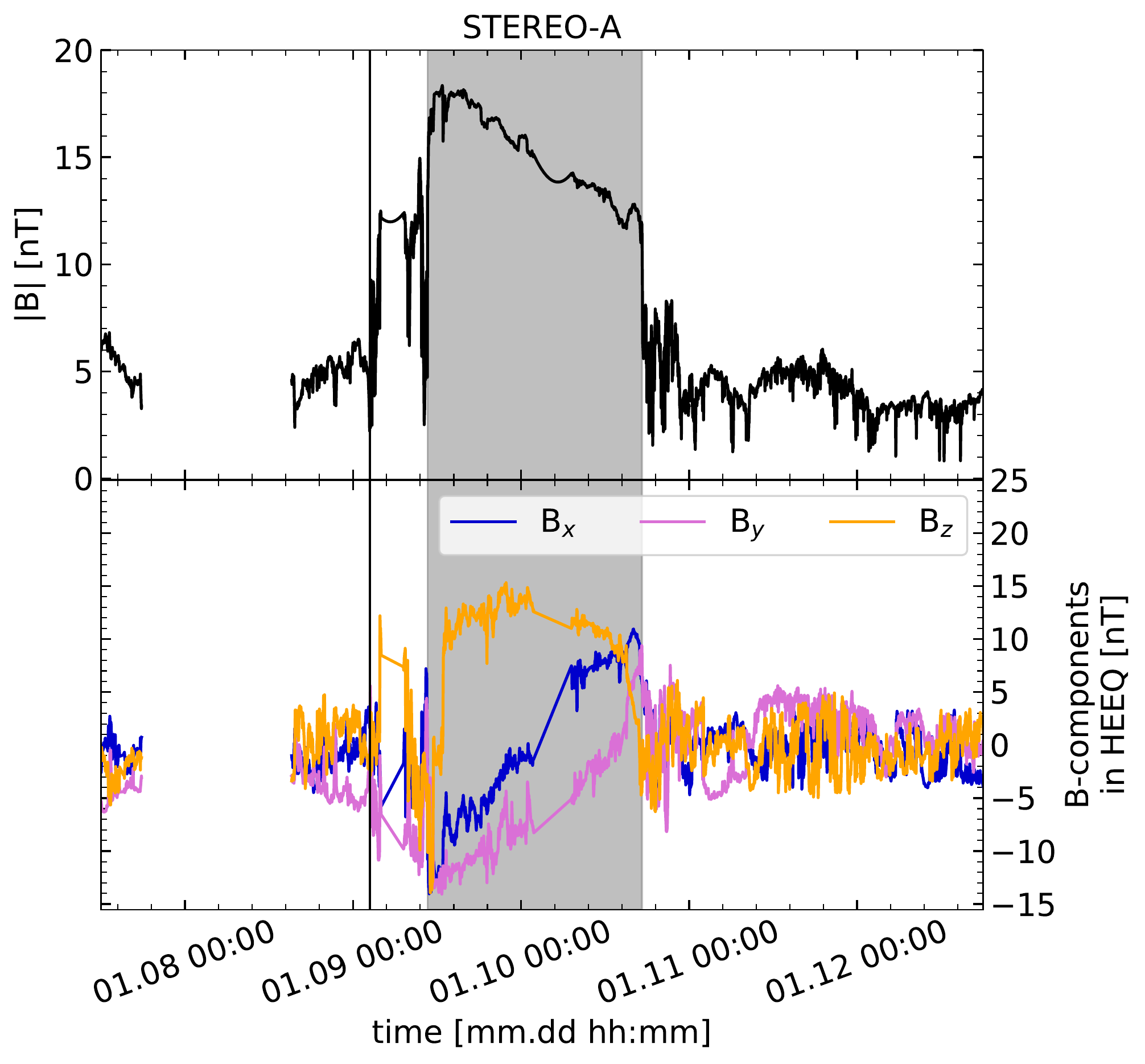}
    \end{minipage}
    \caption{Magnetic field measurements (absolute upper panels; Bx, By, and Bz lower panels) by  Venus Express (top 2 panels) and STERO-A (bottom 2 panels) around the time of the CME arrival at those spacecraft. The in situ shock arrival is marked with a vertical black line and the flux rope signatures at Venus Express and STEREO-A are shown in grey-shaded regions. Data gaps in the Venus Express time--series reflect magnetospheric crossings that have been removed. The spike half way during the flux rope passage from Venus Express is associated with the Venus magnetosheath crossing that was not fully filtered out in the provided data.
              }
 \label{insitufluxrope}
 \end{figure}

The in situ signatures indicate the presence of a sheath region that precedes a long-lasting flux rope featuring a clear magnetic field rotation (see Figure \ref{insitufluxrope}). In Table \ref{table:1} we provide the shock arrival time $t_s$, the leading edge time $t_{f start}$, and trailing edge time $t_{f end}$ of the magnetic cloud as defined by \cite{good_self-similarity_2019}, who also over--plotted the signatures at the two spacecraft and conclude that they match rather well. This suggests that the general magnetic cloud structure did not significantly evolve  during its propagation from one spacecraft to the other. This is also qualitatively depicted in Figure \ref{insitufluxrope}: For both spacecraft, the magnetic field rotation has overall similar patterns; the $B_x$ and $B_y$ components rotate from negative to positive, while the $B_z$ component maintains primarily a northward orientation throughout the flux rope. The magnetic cloud duration $\Delta t$, also given in Table \ref{table:1}, increased by slightly over 2 hours between the two spacecraft, which is likely due to expansion.


\begin{table*}
\caption{ICME shock arrival time $t_{s}$ as well as flux rope start $t_{f start}$ and end $t_{f end}$ times, and duration of the flux rope passage $\Delta$t for Venus Express and STEREO-A spacecraft (SC). These timestamps are taken from \cite{good_self-similarity_2019}.}
\label{table:1}      
\centering                          
\begin{tabular}{c c c c c} 
\hline\hline                 
SC & $t_{s}$ & $t_{f start}$ & $t_{f end}$ & $\Delta t$ \\  
\hline                        
   Venus Express & 8 January 2013 09:22 & 8 January 2013 15:24 & 9 January 2013 19:48 & 28h14min \\[3pt]
   STEREO-A & 9 January 2013 02:25 & 9 January 2013 10:39 & 10 January 2013 17:17 & 30h38min \\[3pt]
\hline                                   
\end{tabular}
\end{table*}


\begin{figure}
   \centering
   \includegraphics[width=0.7\hsize]{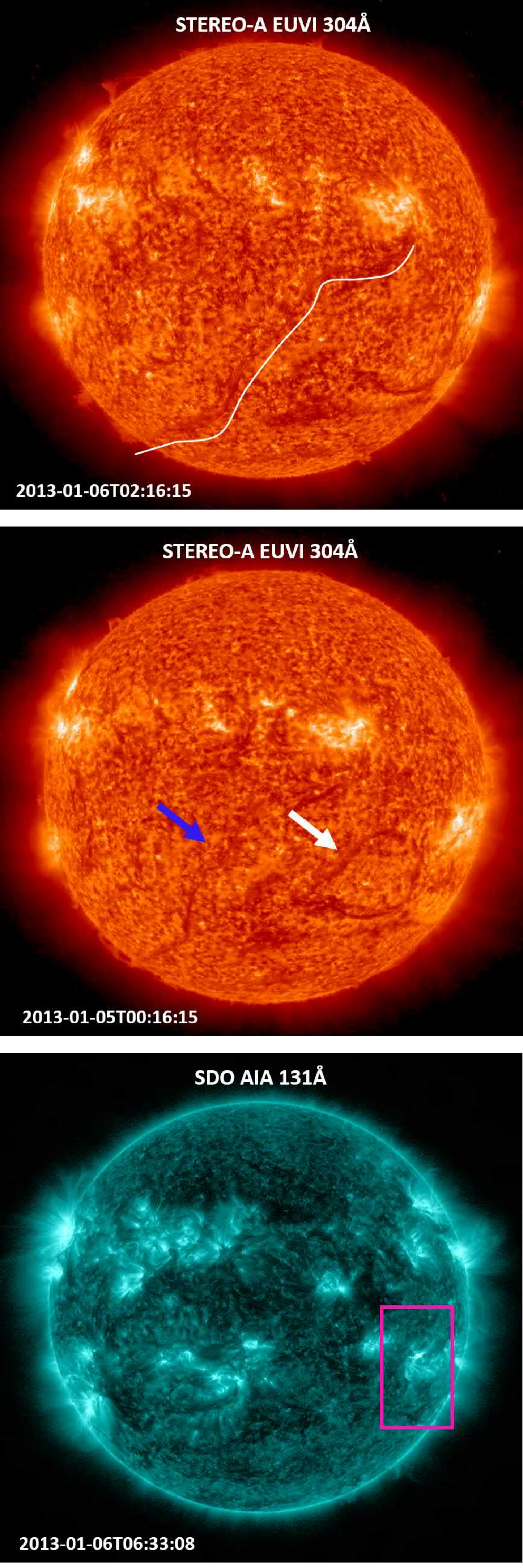}
      \caption{Top image: Filament, indicated by a thin white (hand-drawn) line, approximately one hour before it erupted, as seen in STEREO-A EUVI filtergrams at 304{\AA}. The white line is placed immediately below the actual structure so that it does not cover it. As it appears the filament was a rather large structure, spanning from north equatorial latitudes to south polar latitudes, with a anticlockwise tilt with respect to the solar equatorial plane. A part of the filament possibly wrapped beyond the instrument's field of view near the south pole. Middle image: STEREO-A EUVI filtergram at 304{\AA} showing both filaments that later on erupted one day apart, producing two distinct CMEs, the primary one indicated by a blue arrow and the earlier erupting one by a white arrow. Bottom image: SDO-AIA filtergrams at 131{\AA} with a magenta rectangle enclosing the location of the third eruption, a solar flare, that produced a CME which was visible in the STEREO-A COR2 white-light images indicated in Figure \ref{coronagraphs} (yellow arrow).
      }
         \label{fil_bef_eru}
\end{figure}


\begin{figure*}[h!]
   \centering
   \includegraphics[trim=0 50 0 50, clip, width=0.95\textwidth]{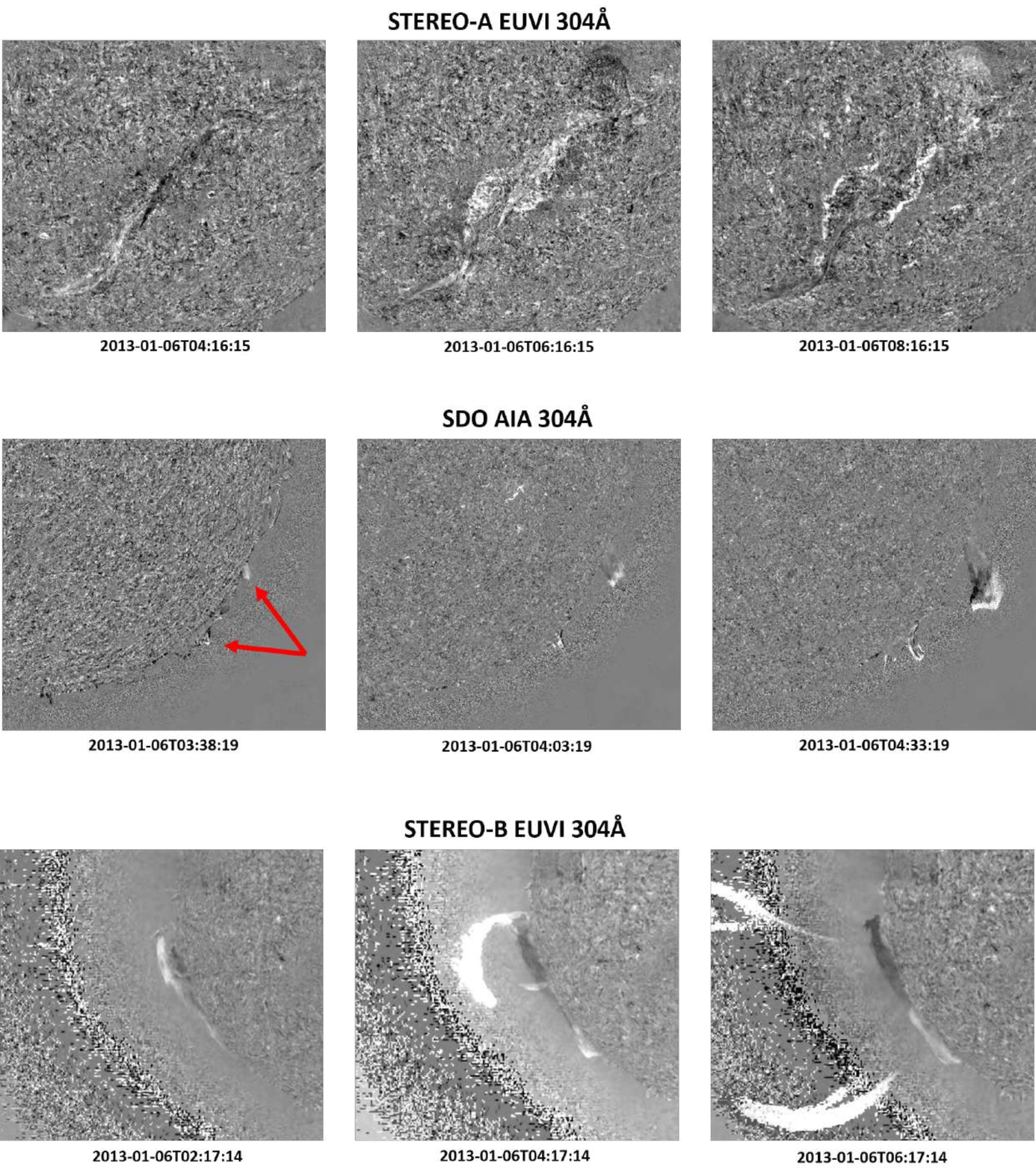}
      \caption{Running-difference images based on EUV filtergrams at 304{\AA} from the EUVI instrument on board STEREO-A (top row) and the AIA instrument on board SDO (middle row), and base-difference images using the filtergrams from the EUVI instrument on board STEREO-B (bottom row). The date format in these images is yyyy-mm-ddThh:mm:ss. The images capture the erupting filament as it evolves in the low corona from three vantage points. For STEREO-A, the filament eruption was located in the field of view of EUVI, and so the erupting filament, the footpoints, and the flare ribbons are clearly visible (top row). For SDO and STEREO-B, the eruption appeared at the limb. In the higher-cadence images of the SDO AIA instrument (middle row), the early signatures of the eruption appeared at the limb at around 03:37:43UT on 2013-01-06 (traces marked with red arrows). These SDO images provide a better estimate of the eruption time. In the STEREO-B field of view, the rising filament material indicative of a flux rope and its legs in the low corona are well captured (bottom row).
              }
        \setlength{\belowcaptionskip}{10pt}
         \label{304filtergrams_rd}
\end{figure*}


\begin{figure*}[h!]
    \centering
    \includegraphics[trim=0 70 0 70, clip, width=0.95\hsize]{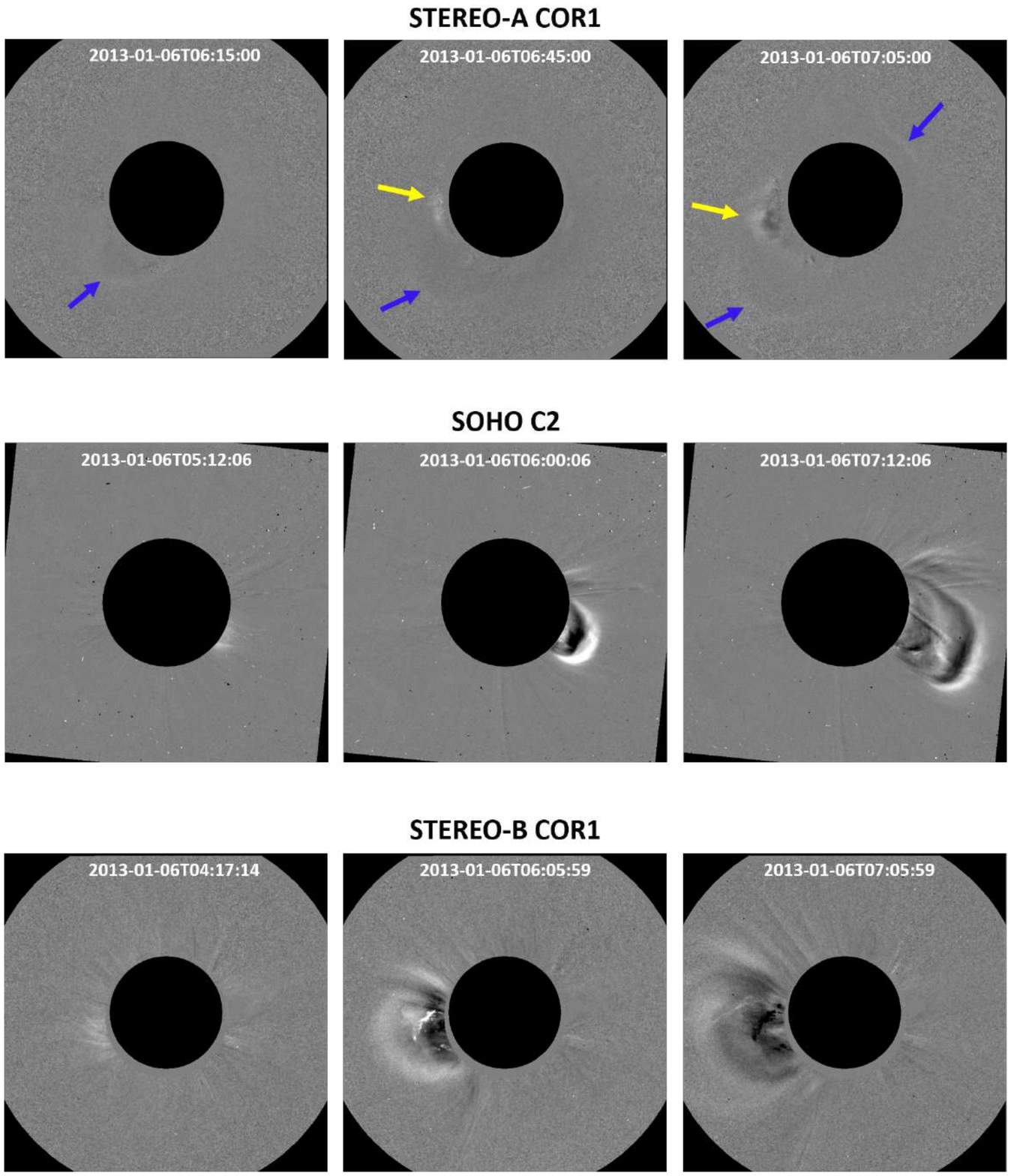}
      \caption{White-light images showing the early evolution of the CME from three different vantage points, namely the two STEREO spacecraft and SOHO (top row: STEREO-A, middle row: SOHO, bottom row: STEREO-B). The date format in these images is yyyy-mm-ddThh:mm:ss. All spacecraft remote observations indicate that the CME propagation was mainly directed towards the south. For STEREO-A COR2, the first signatures of the CME appear to the southeast, as marked with the blue arrow. Only about an hour later did the northern signature appear in the COR2 field of view (also marked in blue). A second eruption appeared in the instrument field of view (marked with yellow arrows), whose source has been identified to be an eruptive flare that happened on the west limb in the field of view of SDO (see Figure \ref{fil_bef_eru}). 
              }
    \label{coronagraphs}
\end{figure*}


\section{Remote-sensing observations}\label{remotesensing}

The source of the CME was an eruption of a large filament that took place from the southern hemisphere on the far side of the Sun from the Earth’s viewpoint. STEREO-A being ${\sim}130^{\circ}$ from Earth (see Figure \ref{orbit} for the spacecraft position) had, however, an approximately head-on view of the source region. The filament extended from the equator to the south-eastern limb and was tilted anticlockwise from the equatorial direction. The filament is indicated by a thin white line in the STEREO - A EUVI 304{\AA} filtergram shown in the top image of Figure \ref{fil_bef_eru}. The same image indicates the part of the filament that is possibly extended beyond the south-east limb.

The erupting material was well observed by the EUV and white-light coronagraph instruments on board STEREO-A, STEREO-B, and SDO. Running-difference images created using the 304{\AA} filtergrams from STEREO-A EUVI and SDO AIA, as well as base-difference images of the 304{\AA} filtergrams from STEREO-B EUVI, are given in Figure \ref{304filtergrams_rd}. For STEREO-B, base-difference images are shown due to a  better contrast. In the STEREO-A field of view, one can clearly see the filament splitting about the polarity inversion line (PIL), the flare ribbons, and the footpoints of the flux rope (top row images in Figure \ref{304filtergrams_rd}). Analysing the AIA difference images provided in the middle row of the same figure, we can deduce that the eruption occurred approximately between 03:31UT and 03:38UT on 6 January 2013 when the first lift-off material appeared off the south-west limb (signatures indicated by red arrows in the first panel of the middle row in Figure \ref{304filtergrams_rd}). In STEREO-B EUVI base-difference images, given in the bottom row of Figure \ref{304filtergrams_rd}, the eruption can only be seen after it rose above $1 \,R_\odot$, but both legs of the rising flux rope are well identifiable.

White-light images taken by LASCO C2 show an increase in intensity at around 05:00UT, and by 06:00UT a bright front surrounding a cavity appeared in the field of view of the instrument (Figure \ref{coronagraphs}, middle row images). The first signatures of the CME in the LASCO C2 field of view suggest that the CME was propagating mostly southward. From the STEREO-A perspective (top row images in Figure \ref{coronagraphs}), the CME first appears off the south-east limb and only about an hour later (at approximately 07:05UT) it appears off the north-west limb. This could be explained by a structure exhibiting a similar extent and tilt as the filament source. The signatures off the south-east are faint yet visible in the top row of Figure \ref{coronagraphs} indicated by blue arrows. Out of all three vantage points investigated, STEREO-B had the clearest view of the part of the CME that propagated radially and close to the equator. This corresponds to the faint structure that later appeared in the north-west in the STEREO-A field of view.

One day prior to the eruption of the extended filament discussed above, a neighbouring filament with a fairly similar inclination and only a few degrees away towards the west limb in the STEREO-A EUVI field of view also erupted. The two filaments are indicated in the middle image in Figure \ref{fil_bef_eru} where the main filament under study is indicated by a blue arrow and the filament that erupted the previous day is shown with a white arrow. This eruption took place between 04:00--06:00UT on 5 January 2013 and it appeared as a narrow CME in coronagraph images. From assessing the EUV and coronagraph images as well as the in situ signatures and applying a GCS analysis to both eruptions, it was deduced that the two filament eruptions did not interact. As it is discussed in Appendix \ref{cone_EUHFORIA}, we also performed a cone model EUHFORIA run to confirm this.

A third eruption was captured in coronagraph images by STEREO-A, which first appeared in the field of view of COR1 on 6 January 2013 at 06:45UT (indicated by yellow arrows in the top row of Figure \ref{coronagraphs}). This third CME originated from a weak eruption at the west limb in the field of view of SDO (within the magenta square in the bottom image of Figure \ref{fil_bef_eru}), and it is a back-sided event from the perspectives of Venus Express and STEREO-A. The EUV signature in SDO imagery also indicates a weak event. Based on the position of the spacecraft (Figure \ref{orbit}), it is expected that this eruption was headed approximately towards Earth and was therefore not captured by Venus Express or STEREO-A. No traces of this CME were visible from the coronagraphs on board STEREO-B or SOHO.


\section{CME properties} \label{properties}

\subsection{Magnetic flux and helicity sign} \label{MF_and_helicity}

Since the eruption was a back-side event from Earth's perspective, there are no on-disc magnetic field observations from the time of the eruption. It is thus not possible to extract information for the magnetic flux of the CME using co-temporal magnetogram-based approaches \citep[e.g. see][]{dissauer_detection_2018, dissauer_statistics_2018, pal_dependence_2018, dissauer_statistics_2019, sarkar_observationally_2020}. In this case, a base value of $80.0\times10^{12}$~Wb was used for the EUHFORIA runs, which represents the toroidal flux and is related to the magnetic field strength via equation 7 given in \citet{verbeke_evolution_2019}. Similarly, base values were used for the uniform density and the temperature of the CME. These are $1\times10^{18}$~kg/m$^3$ and $0.8\times10^{6}$~K, respectively \citep{verbeke_evolution_2019}.

\begin{figure*}
   \centering
   \includegraphics[width=0.85\hsize]{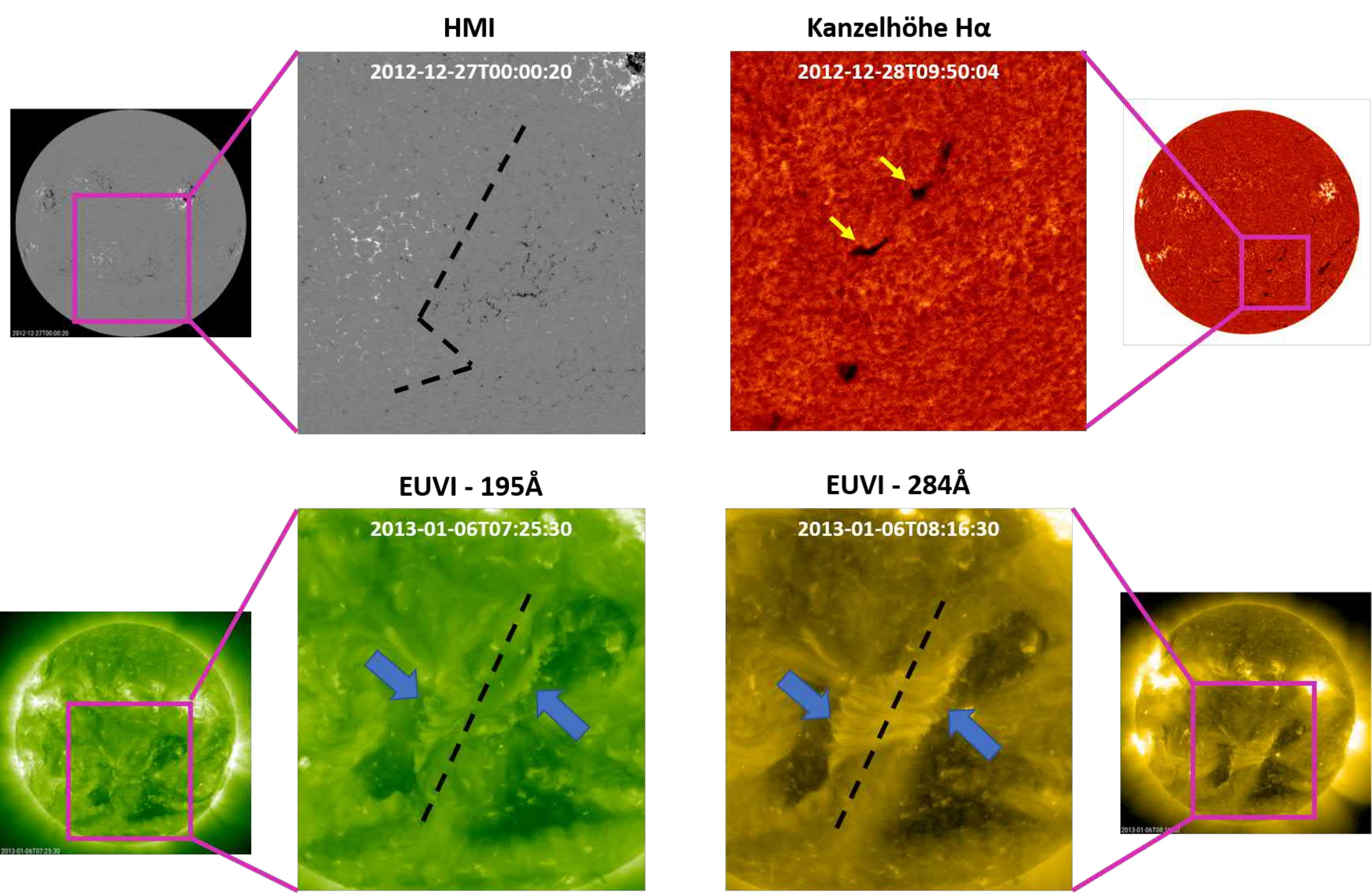}
      \caption{Filament and post-eruption characteristics used in identifying the flux rope helicity sign. The date format in these images is yyyy-mm-ddThh:mm:ss. The magenta squares show the area of the Sun that is being magnified. The images in the top left corner show the magnetic field topology captured by the HMI magnetogram a few days prior to the eruption. The black line marks the PIL. In the top right corner there are H$\alpha$ images of the Sun taken with the solar telescope at Kanzelh{\"o}he Solar Observatory and compensated for limb-darkening following the method by \citet{chatzistergos_analysis_2018}. Barb structures are indicated with yellow arrows. The STEREO-A EUVI filtergrams at 195{\AA} and 284{\AA} correspond to the left and right images in the bottom row, respectively, and they show the post-eruption arcade and flare ribbons. The latter is indicated by blue arrows in the bottom row zoomed images, while the black dashed line indicates the PIL.
              }
         \label{helicity_analysis}
\end{figure*}

To determine the helicity sign of the magnetic field structure of the CME, we investigated observational proxies for determining the helicity sign using remote-sensing imaging observations \citep{chen_imaging_2014, palmerio_determining_2017,ouyang_chirality_2017}. The filament was visible for several days before the eruption. On 27 December 2012, it was within the field of view of Earth and, therefore, we used SDO/HMI magnetograms to analyse its magnetic field topology. The photospheric magnetic field is positive to negative from the solar east to the solar west, with the filament lying above the PIL  (top left panel in Figure \ref{helicity_analysis}). We note that H$\alpha$ images (Figure \ref{helicity_analysis}, top right corner) show that the orientation of the filament barbs (marked with arrows) relative to the filament axis along the PIL indicate a sinistral, positive helicity, flux rope. STEREO-A EUV images (bottom panels) of the post eruption arcade (PEA) show the flaring arcades being right-skewed in comparison to the underlying magnetic PIL. This supports the positive helicity sign indicated by the pre-eruption signatures of the filament.


\subsection{Morphology and kinematics} \label{mor_kin}

\begin{figure*}
   \centering
   \includegraphics[width=0.85\hsize]{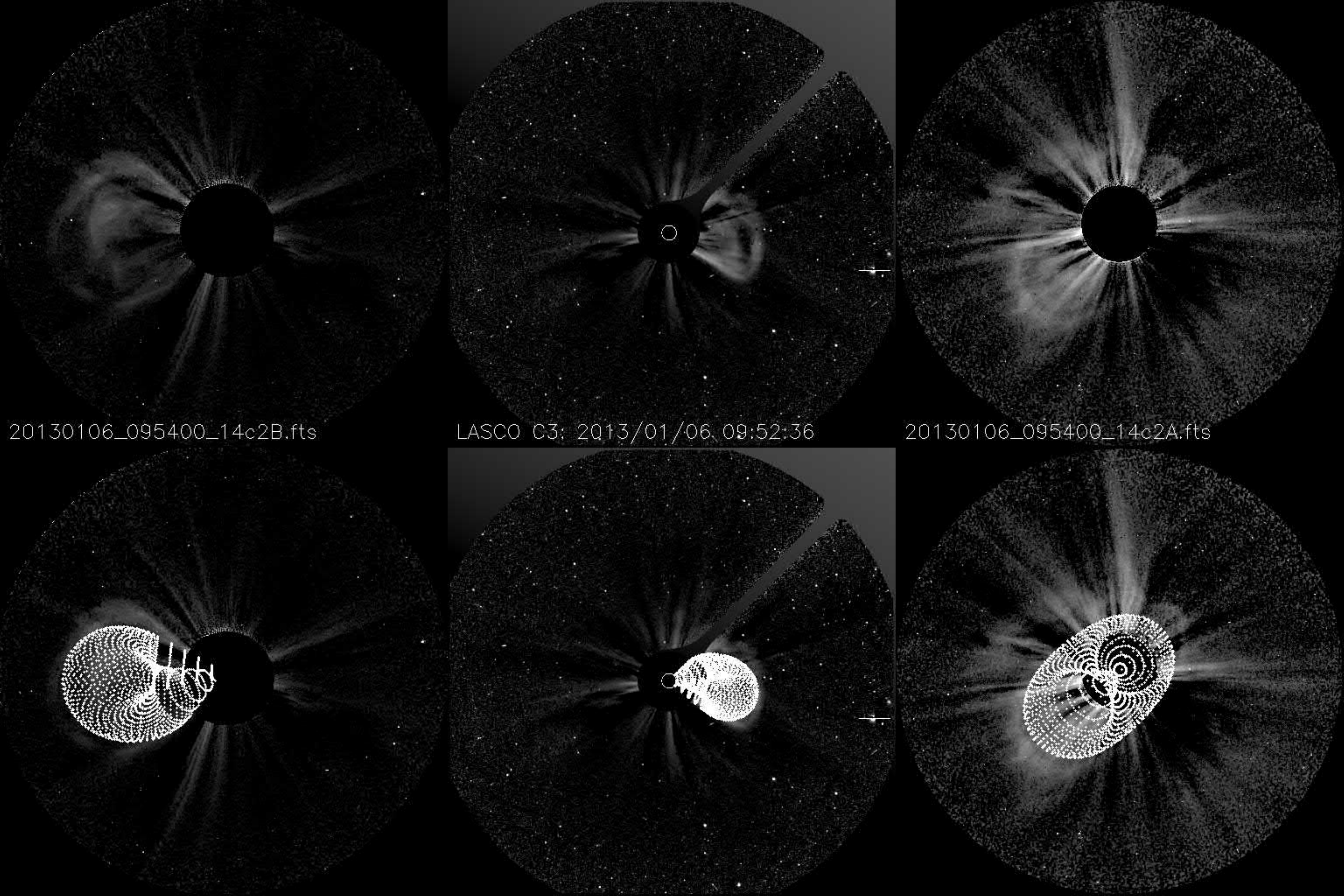}
      \caption{Example of the GCS fitting applied to the CME structure seen in white-light images by both STEREO spacecraft (left column for B and right column for A) and SOHO (middle). The fitting was focused on enclosing the cavity and the flux rope of the CME, but not the shock structure.
              }
         \label{GCS_fitting}
\end{figure*}

\begin{figure}
\centering
\resizebox{0.8\hsize}{!}
            {\includegraphics[]{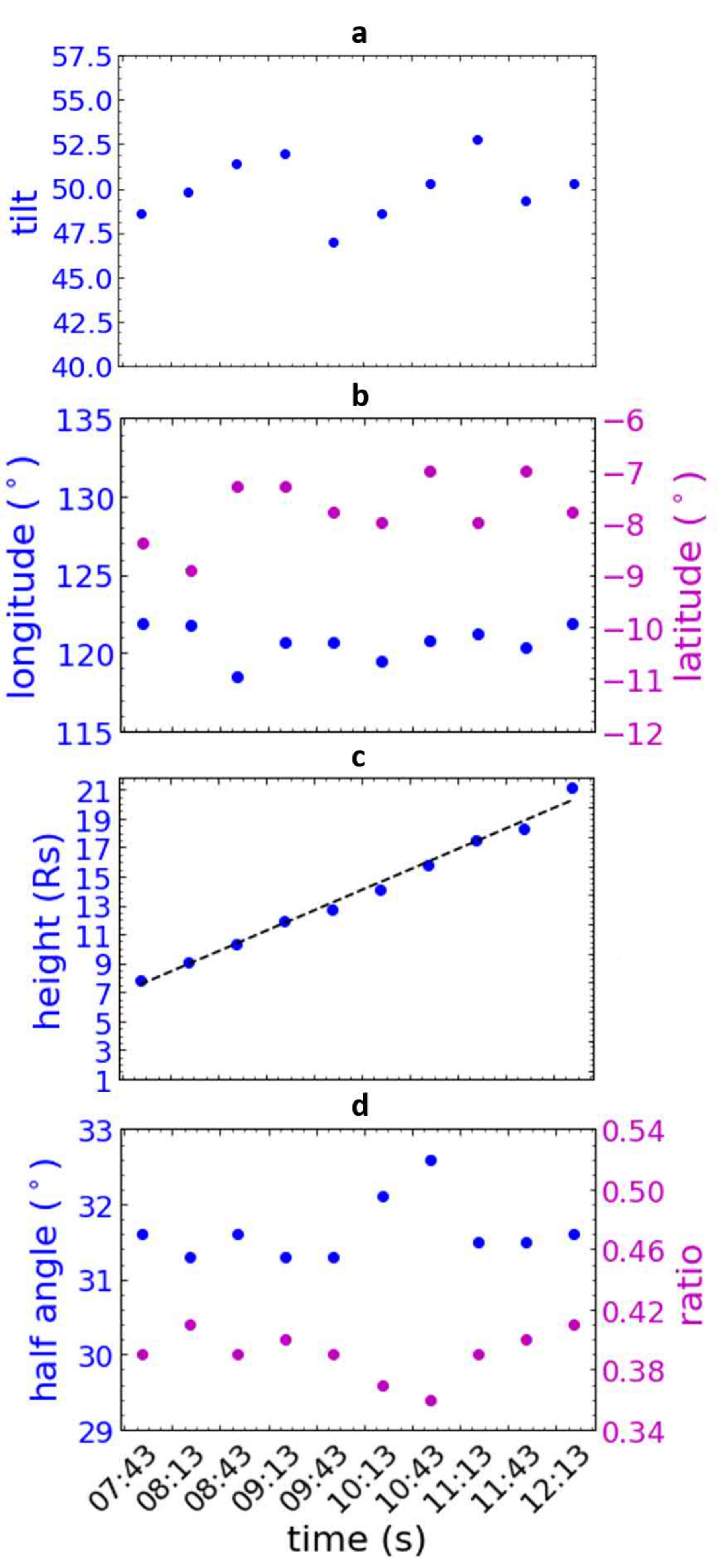}}
      \caption{Scatter plots of the CME morphological parameters extracted using the GCS. The methodology was applied at different time steps to investigate whether the CME underwent deflection low in the corona, or other kinematic and/or morphological changes.
              }
         \label{GCS_scatter}
   \end{figure}

We used a forward modelling approach employing the GCS geometric model to determine the size, kinematics, and propagation direction of the CME \citep{thernisien_modeling_2006, thernisien_forward_2009}. We performed the analysis using white-light images taken at ten time steps 30 minutes apart, thus providing an estimate of the evolution of the CME in the corona. An example of one of the GCS fittings is given in Figure \ref{GCS_fitting}. At all times, an attempt to trace the same features was made, for example, focusing on always enclosing the cavity with the fitting mesh and avoiding the outermost brightening that most likely corresponds to the shock. The resulting evolution of the geometric parameters from all the fittings is shown in the scatter plots in Figure \ref{GCS_scatter}. As it can be seen, there is only little variability of the fitted parameters. The CME speed and the time when its apex reached 21.5$R_\odot$ (i.e. corresponding to the inner boundary of EUHFORIA’s heliospheric model) were determined by using the least square fitting applied to the CME heights obtained with GCS reconstructions from the ten time stamps (see panel c of Figure \ref{GCS_scatter}). The estimated speed is 571 km/s and the date--time of the CME passage at 21.5$R_\odot$ is 6 January 2013 12:49:33 UT. These values are also listed in rows 6 and 7 of Table \ref{table:2}.

The small fluctuations in the time series of longitude, latitude, tilt, aspect ratio, and half angle visible in Figure \ref{GCS_scatter} reflect uncertainties in the fitting process, rather than indicating
actual changes in the CME features (position and shape). These uncertainties can arise from the following: (1) a lack of unambiguously clear similar structures in the images, which is a consequence of the nature of the CME as well as the Thomson scattering and the separation of the spacecraft; (2) shoehorning of the GCS shape; (3) difficulty in precisely fitting the GCS mesh to the white-light image structures and in subjectivity in the fitting; and (4) sensitivities to image quality and vantage points. Thus, the output parameters of a GCS fitting are only rough estimates of the actual location and geometry of the CME. We note that the small fluctuations would incur only minor changes to the heliopsheric model results as they are, for example, smaller than the variation in the CME input parameter range employed by \cite{mays_ensemble_2015} in their sensitivity analysis of the WSA-Enlil Cone model. Thus, we consider, in the following, only the average value for each parameter. These values are given in Table \ref{table:2}.


\begin{table}
\caption{CME morphological and kinematic parameters as extracted using the GCS geometric reconstruction method. The longitude and latitude are given in the HEliospheric EQuatorial (HEEQ) coordinate system. The tilt is determined relative to the solar equator.}
\label{table:2}      
\centering                          
\begin{tabular}{c | c}      
HEEQ longitude ($\phi$) & 120.7$^{\circ}$ \\[5pt]
HEEQ latitude ($\theta$) & -7.8$^{\circ}$ \\[5pt]
tilt ($\gamma$) & 50.0$^{\circ}$ \\[5pt]
half angle ($\alpha$) & 31.6$^{\circ}$ \\[5pt]
aspect ratio ($\kappa$) & 0.39 \\[5pt]
21.5$R_\odot$ arrival & 6 January 2013 12:49:33 UT \\[5pt]
speed ($u$) & 571 km/s \\[5pt]
Edge--on radius ($R_{\mathrm{\omega_{EO}}}$) & 8.4$R_\odot$ \\ [5pt]
Face--on radius ($R_{\mathrm{\omega_{FO}}}$) & 17.5$R_\odot$ \\ [5pt]
Averaged radius ($\langle R \rangle$) & 13.3$R_\odot$ \\
\end{tabular}
\end{table}


\subsection{Constraining the Spheromak radius}\label{GCS_spher}

When performing the heliospheric magnetohydrodynamic (MHD) simulations (Section \ref{model}), we used a spheromak CME model that is characterized by a magnetic field configuration filling a spherical volume. The temperature and the density inside the spheromak-CME are constant. As these are difficult to determine from remote-sensing observations, in EUHFORIA for simplicity we have set a default value for both, which are given in Table \ref{table3}. The same values are used for the cone and the spheromak CME representations, which are currently implemented in EUHFORIA, and they were first introduced by \citet{pomoell_euhforia_2018} for the cone model and maintained as the default values in \citet{verbeke_evolution_2019} for the spheromak model implementation. With respect to the ambient solar wind, the fast solar wind plasma thermal pressure was set to be 3.3nPa, which is equivalent to a plasma temperature for the fast solar wind of 0.8MK. The spheromak temperature is equal to that value of the fast solar wind temperature. Regarding the plasma density, the spheromak is indeed significantly less dense than the ambient slow solar wind and the shock and sheath formed ahead of the spheromak. This is consistent with the observational analysis in \citet{temmer_deriving_2021}, where from observations at 1AU, a rather nice linear relation was derived with a factor of about 2 to 3 between ambient solar wind density 24 hours ahead of the CME and sheath density, independent of the CME speed.  Thus, the spherical volume of the spheromak is a low-density, high-temperature cavity. It is important at this stage to define the radius of the spheromak based on the GCS fittings of a crescent shaped structure, two shapes that are essentially different but not totally incompatible.

The GCS provides two dimensions of the CME, namely the edge-on angular width, $\omega_{EO}$, and the face-on angular width, $\omega_{FO}$, given by:

    \begin{equation}
      \omega_{EO}= 2\delta = \arcsin{\kappa} \,,
   \end{equation}

   \begin{equation}
      \omega_{FO} = 2(\alpha + \delta)\,,
   \end{equation}

where $\kappa$ is the GCS ratio and $\alpha$ is the GCS half angle \citep{thernisien_implementation_2011}. In this study, the $\omega_{EO}$ and $\omega_{FO}$ values are used to define three different radii for the spheromak, given by:

        \begin{equation}
         R_{\mathrm{\omega_{EO}}} = 21.5\sin{(\omega_{EO}/2)} \,,
        \end{equation}
        \begin{equation}
         R_{\mathrm{\omega_{FO}}} = 21.5\sin{(\omega_{FO}/2)} \,,
        \end{equation}
        \begin{equation}
         \langle R \rangle = 21.5\sin{((\omega_{EO}+\omega_{FO})/4)} \,.\end{equation}

The radii obtained are thus $R_{\mathrm{\omega_{EO}}} = 8.4R_\odot$, $R_{\mathrm{\omega_{FO}}} = 17.5R_\odot$, and $\langle R \rangle = 13.3R_\odot$. All the CME input parameters used for the heliospheric MHD simulation runs are summarised in Table \ref{table3}.



 \section{EUHFORIA model}\label{model}
 
 \subsection{Architecture}\label{architec}
 EUHFORIA \citep{pomoell_euhforia_2018} is a recently developed MHD model that aims to reproduce the spatial and temporal evolution of CMEs throughout the inner heliosphere up to 2 AU, propagating in realistic ambient solar wind conditions. It is a data-driven model consisting of two building blocks, the 'coronal domain' and the 'inner heliosphere domain'. The coronal domain stretches from the photosphere up to 0.1 AU, and the inner heliosphere domain starts at 0.1 AU and has an outer boundary set at 2 AU. For the coronal domain, EUHFORIA employs a two-part magnetic field model, the potential field source surface \citep[PFSS; ][]{altschuler_magnetic_1969} for the lower corona, and the Schatten current sheet \citep[SCS; ][]{schatten_model_1969} model for the upper corona, as well as an empirical solar wind model based on the Wang-Sheeley-Arge \citep[WSA; ][]{arge_improved_2003} model. These models produce magnetic field and plasma conditions at 0.1 AU, which act as boundary conditions for the inner heliospheric model, and aim to realistically describe the large-scale solar wind streams at the 0.1 AU boundary. An important input parameter for the coronal model are synoptic magnetograms provided by GONG. In this study, we employed the recently updated ADAPT magnetograms \citep{arge_air_2010}. For the inner heliosphere domain, the CME interaction with the ambient solar wind and, subsequently, its evolution and propagation are modelled in a self-similar manner by solving the time-dependent MHD equations in three-dimensional space.


CMEs were inserted in the simulation at the inner boundary of the inner heliosphere domain at 0.1 AU. Two principal CME models were implemented in EUHFORIA, the cone CME model \cite[similar to][]{odstrcil_three-dimensional_1999} and the spheromak CME model, similar to the one described in  \citet{verbeke_evolution_2019}, for example. The cone CME model considers the CME only as a hydrodynamic structure with a constant speed and angular width propagating in a constant direction. It thus does not trace the evolution of its magnetic structure. The spheromak CME model considers the CME as a closed spherical bubble enclosing an axisymmetric twisted magnetic field. It assumes a linear force-free configuration with a constant internal density and temperature. The spheromak was introduced into the domain via a time-dependent boundary condition at 0.1 AU. During the time the spheromak emerges into the domain (related to the speed at which the structure propagates), the structure is connected to the boundary. However, after it has been fully emerged, the flux rope does not connect to the inner boundary. It is then fully detached. Since the spheromak is contained within a spherical volume, the magnetic field structure does not have legs that would attach it to the lower coronal domain. We refer the reader to \citet{verbeke_evolution_2019} for further details regarding the implementation. The spheromak expansion is due to multiple effects, including the reaction to the change of the thermal pressure in the ambient solar wind, as the spheromak moves away from the Sun. The spheromak expands to reach plasma and magnetic pressure balance with the ambient solar wind. However, this never happens as the CME continues to propagate outwards, so the external pressure keeps decreasing. We note that when the spheromak is inserted, the structure is in general not in (total) pressure balance with its surroundings. For more details, see \citet{scolini_observation-based_2019}.

For the heliospheric domain, we used a uniform grid in all directions. The spatial resolution of the MHD heliospheric domain is 4 degrees in longitude and latitude, and a total of 256 cells in the radial direction. This is the default EUHFORIA set up for the heliospheric domain.

\subsection{Output} \label{output}

\subsubsection{Choice of spheromak rotation angle}\label{hel_tilt}

When modelling CMEs using the spheromak in EUHFORIA, an additional parameter beyond those described in previous sections needs to be specified, namely the angle that the axis of symmetry of the magnetic field configuration subtends with respect to the equatorial plane. In the GCS model or other similar loop-like flux rope models, the direction of the axis of the structure with respect to the equatorial plane is given by the tilt angle. However, this tilt angle is related to, but not equivalent to, the spheromak rotation angle. The major difference is that the tilt angle describes both a morphological feature (aspect ratio) and a property of the magnetic field; whereas for the spheromak, it relates only to the structure of the magnetic field. Since the white-light observations do not directly convey information on the magnetic field structure, the internal rotation of the spheromak essentially remains a free parameter in the model.

\begin{figure*}
   \centering
   \includegraphics[width=0.75\hsize]{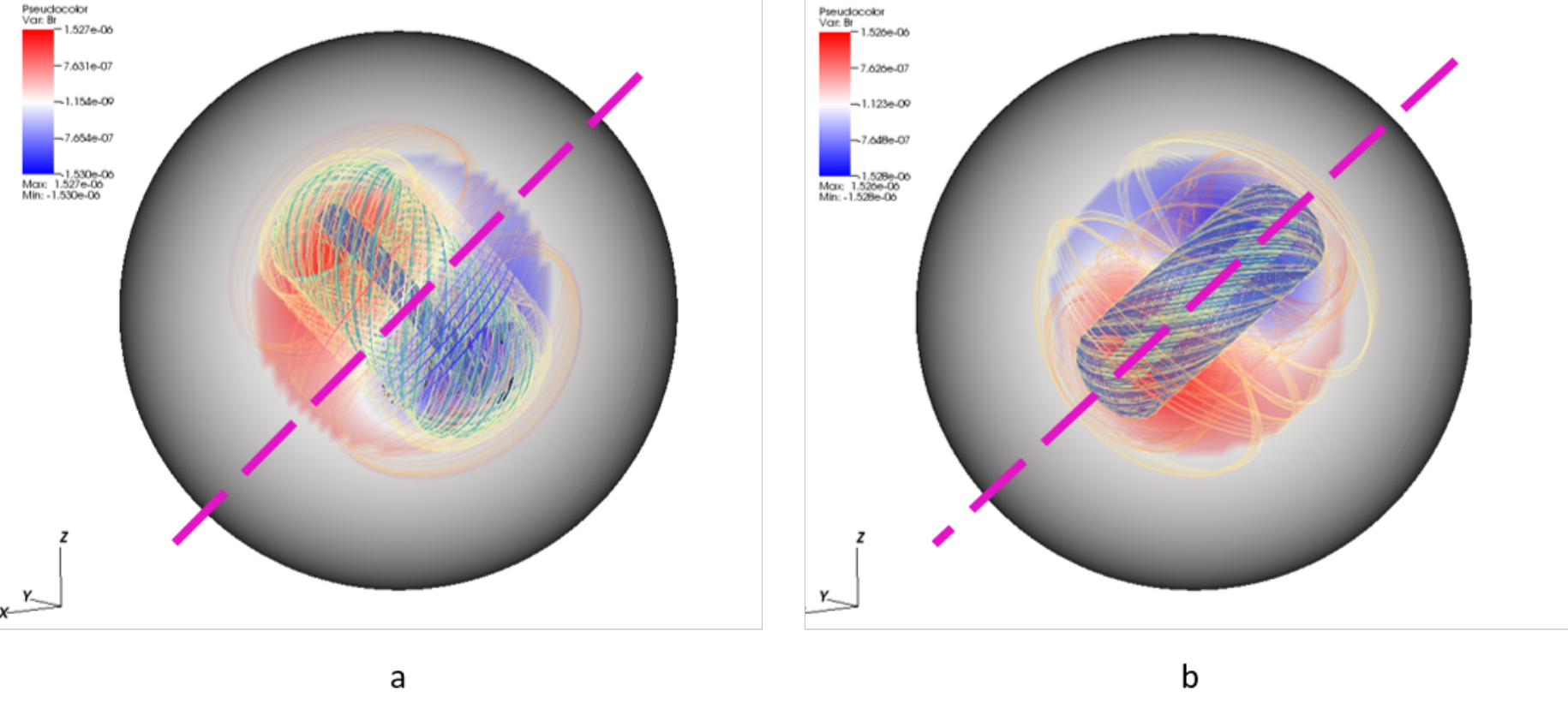}
      \caption{Magnetic field topology of the modelled spheromak CME based on the selected tilt. Left: Magnetic structure for the case when injecting a spheromak with a tilt value obtained from the GCS fitting. Right: Spheromak with rotation angle equal to the tilt increased by 90 degrees anticlockwise. The magenta dashed line marks the orientation of the observed filament and thus the GCS tilt.
              }
         \label{tilt_plus}
   \end{figure*}

Figure \ref{tilt_plus} shows illustrative magnetic field lines for two different orientations of the spheromak together with the direction of the PIL and filament in the corona indicated by the magenta dashed line. In this work, we chose the rotation angle according to panel b. With this choice, the axis of symmetry of the spheromak is perpendicular to the direction of the PIL. The choice is motivated in order to obtain magnetic field rotation (blue field lines) consistent with the observationally inferred magnetic field structure. It is important to note, however, that the maximum of the field strength occurs along the axis of symmetry in the spheromak. We also performed model runs using the rotation indicated in panel a and confirmed that incorrect rotation of the magnetic field vectors was obtained.

With the rotation angle determined, three different EUHFORIA runs were performed modelling the CME as a spheromak. The parameters used are detailed in Table \ref{table3}. Apart from the spheromak radius, all other input parameters were the same among this set of runs. The different values for the radius of the spheromak employed in each run are $R_{\mathrm{\omega_{EO}}}$, $R_{\mathrm{\omega_{FO}}}$, and $\langle R \rangle$, respectively, as derived in Section \ref{GCS_spher}.


\subsubsection{Overview of spheromak model run results}

\begin{sidewaysfigure*}
   \resizebox{1\hsize}{!}
            {\includegraphics[trim=0 100 0 100, clip]{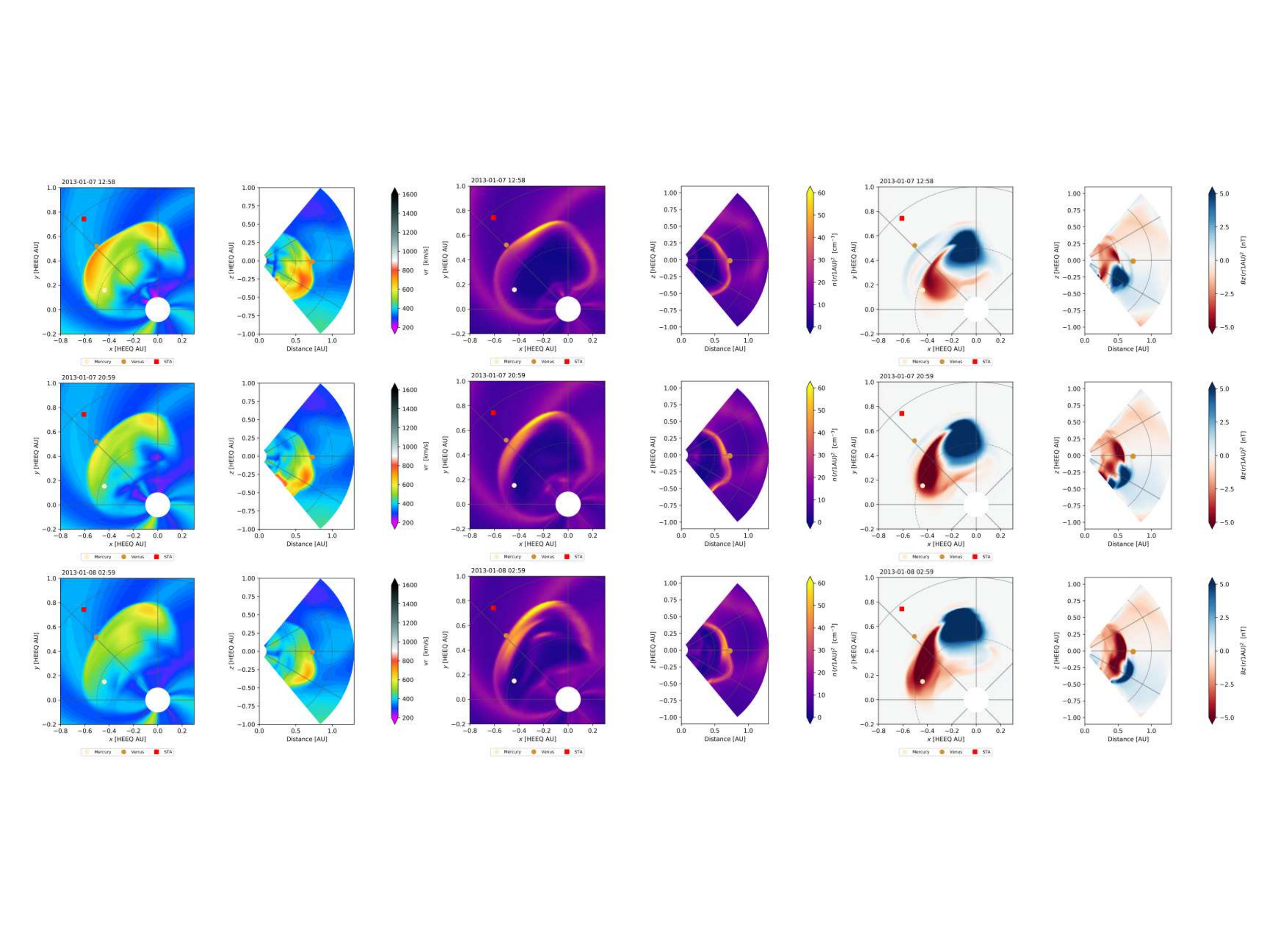}}
      \caption{Spatial profiles of the modelled radial velocity, number density, and HEEQ $B_z$-component of the magnetic field of the CME upon arrival at Venus Express orbit. The top row shows the model output for a spheromak with radius $R_{\mathrm{\omega_{EO}}}$, the middle row for $\langle R \rangle$, and the bottom row for $R_{\mathrm{\omega_{FO}}}$. Both equatorial and meridional cuts are provided to offer a perspective of what portion of the modelled CME structure crossed the spacecraft.
              }
         \label{VEX_eq_mer_minus90}
   \end{sidewaysfigure*}

\begin{sidewaysfigure*}
   \resizebox{1\hsize}{!}
            {\includegraphics[trim=0 100 0 100, clip]{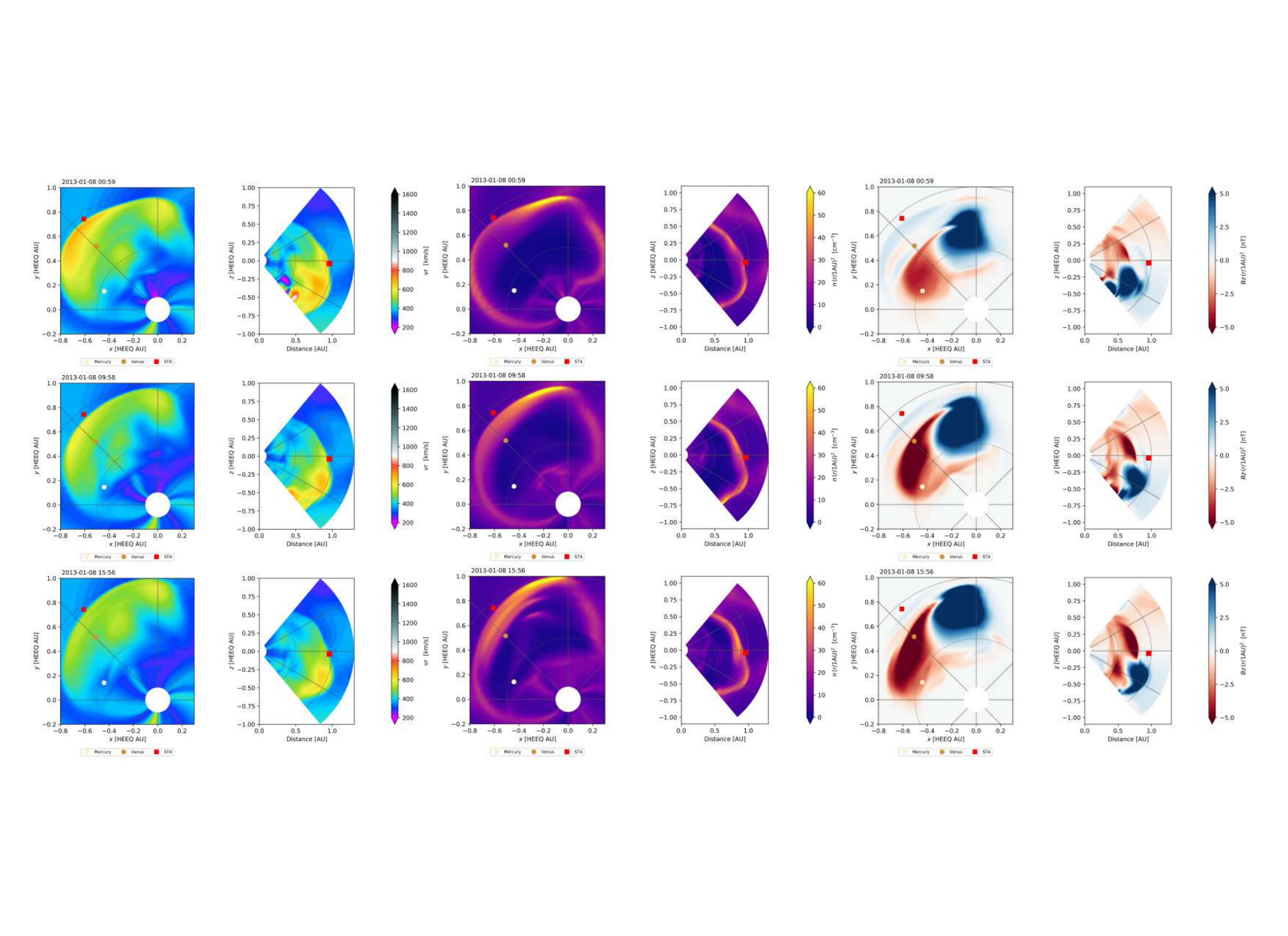}}
      \caption{Meridional and equatorial cuts for the modelled radial velocity, number density, and HEEQ $B_z$-component of the magnetic field of the CME upon arrival at STEREO-A orbit. The order of placement is based on the increasing radius of the spheromak CME from the top to bottom row ($R_{\mathrm{\omega_{EO}}}$, $\langle R \rangle$, and $R_{\mathrm{\omega_{FO}}}$). The format is the same as in Figure \ref{VEX_eq_mer_minus90}.
              }
         \label{STA_eq_mer_minus90}
   \end{sidewaysfigure*}

Figures \ref{VEX_eq_mer_minus90} and \ref{STA_eq_mer_minus90} display the results of the spheromak model runs in the equatorial and meridional cuts at the position of Venus Express and STEREO-A, showing the spatial distribution of the modelled radial velocity, number density, and HEliospheric EQuatorial (HEEQ) $B_z$-component of the magnetic field for the CME at its arrival times at Venus Express and STEREO-A, respectively. We note that both the number density and the $B_z$ component were scaled by $(r/1 AU)^2$ in order to counteract the approximate inverse square variation of the quantities in the solar wind. Each row shows the result for a different CME model run based on the parameters given in table \ref{table3}. Each run is based on a different radius for the spheromak, with the images in the top row showing the results for $R_{\mathrm{\omega_{EO}}} = 8.4R_\odot$, in the middle row for $\langle R \rangle = 13.3R_\odot$, and in the bottom row for $R_{\mathrm{\omega_{FO}}} = 17.5R_\odot$.

As it can be seen in Figures \ref{VEX_eq_mer_minus90} and \ref{STA_eq_mer_minus90}, in the equatorial plane both Venus Express and STEREO-A had a central encounter with the modelled CME for all cases. A near-flank encounter with Mercury, expected from the in situ signatures, is also captured by the model output. In the meridional cuts one can see that Venus Express and STEREO-A were crossing a zone of weak $B_z$-component, with very little rotation.

From the velocity profiles in both Figures \ref{VEX_eq_mer_minus90} and \ref{STA_eq_mer_minus90}, it can be seen that two islands with a higher speed form in the equatorial plane for the smallest size spheromak ($R_{\mathrm{\omega_{EO}}}$, top row, yellow to orange colours). When a CME of a larger radius is modelled (middle and bottom rows), islands of a high speed are less prominent. In the meridional cuts, it can be seen that the fastest portion of the CME is concentrated on the south edge of the structure at approximately –25$^\circ$ latitude. This speed distribution drives the CME to bulge at its southern part. This bulging is more apparent in the largest radius spheromak modelled ($R_{\mathrm{\omega_{FO}}}$, bottom row). This change in the profile of the CME among the runs of different spheromak radii is also apparent for the density and $B_z$-component images. The islands visible in the velocity images correspond to the portions carrying a flux rope of a different polarity (positive -- negative) in the $B_z$-component. The phenomenon is less prominent for the larger spheromak modelled (bottom rows of Figures \ref{VEX_eq_mer_minus90} and \ref{STA_eq_mer_minus90}). The structure of the modelled spheromaks are similar at Venus Express and at STEREO-A orbits. This is in accordance with the observations where the flux rope signatures registered in situ by the two spacecraft do not show the flux rope evolving considerably apart from a slight expansion. In Figure \ref{VEX_B_comp_eq_mer_minus90}, we also provide the meridional and equatorial cuts for the $B_x$ and $B_y$-components of the model outputs at Venus Express, with the smaller radius modelled spheromak ($R_{\mathrm{\omega_{EO}}}$) being shown in the top row, while the averaged sized one ($\langle R \rangle$) is given in the middle row, and the larger one ($R_{\mathrm{\omega_{FO}}}$) in the bottom row.

\subsubsection{Comparing model output to in situ time-series}

\begin{sidewaysfigure*}
   \resizebox{1\hsize}{!}
            {\includegraphics[trim=0 2.2cm 2cm 0, clip]{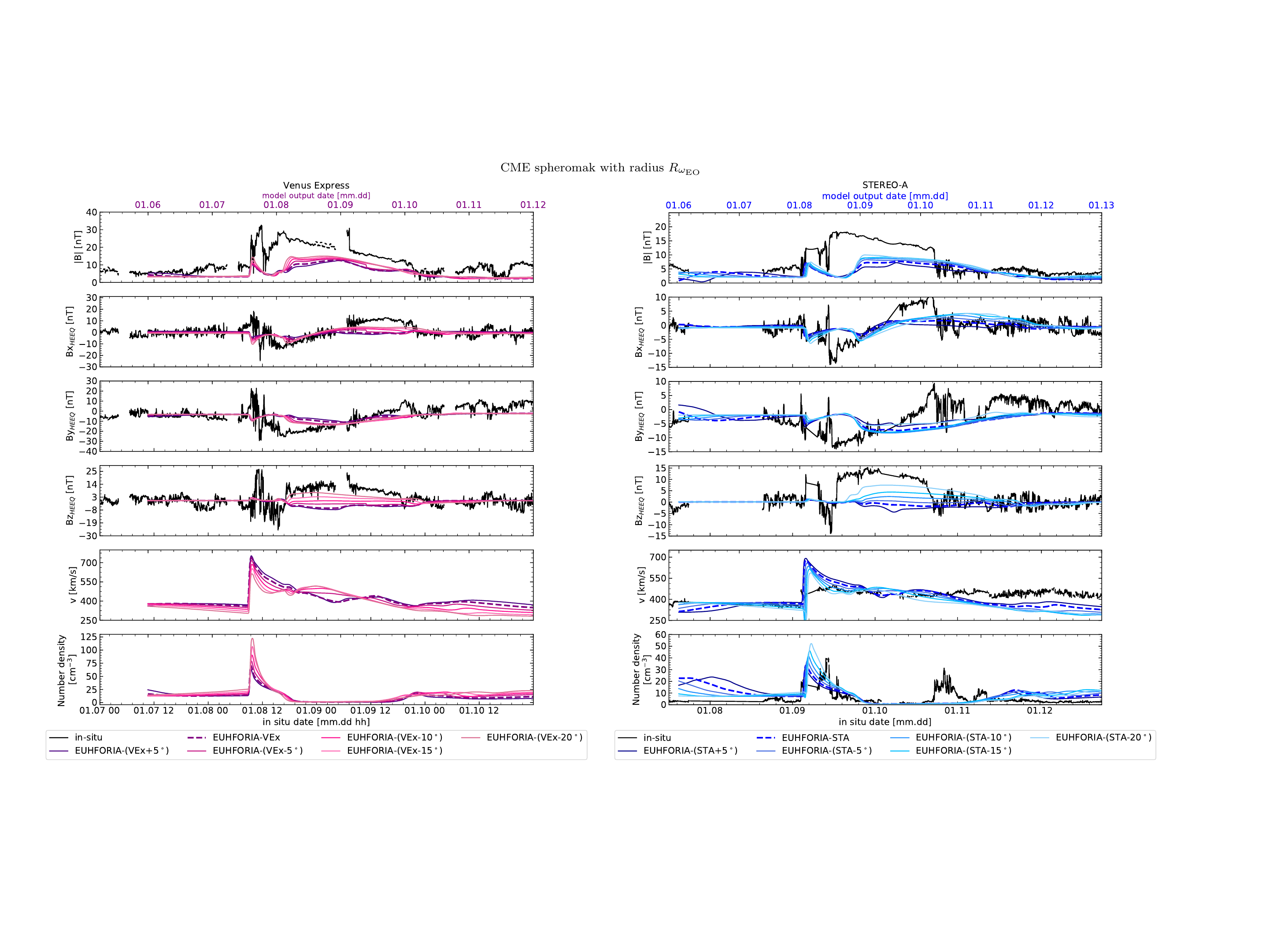}}
      \caption{Time series of the modelled plasma and magnetic field parameters at Venus Express and STEREO-A for a spheromak of size $R_{\mathrm{\omega_{EO}}}$. The in situ observations are given for comparison (black solid lines). The dashed coloured lines indicate the modelled time series at the exact location of the spacecraft, while the solid coloured lines are for the virtual spacecraft. All modelled curves were shifted by 1d18hr for Venus Express and 1d16hr for STEREO-A.
              }
         \label{timeseries_small}
   \end{sidewaysfigure*}

\begin{sidewaysfigure*}
   \resizebox{1\hsize}{!}
            {\includegraphics[trim=0 2.2cm 2cm 0, clip]{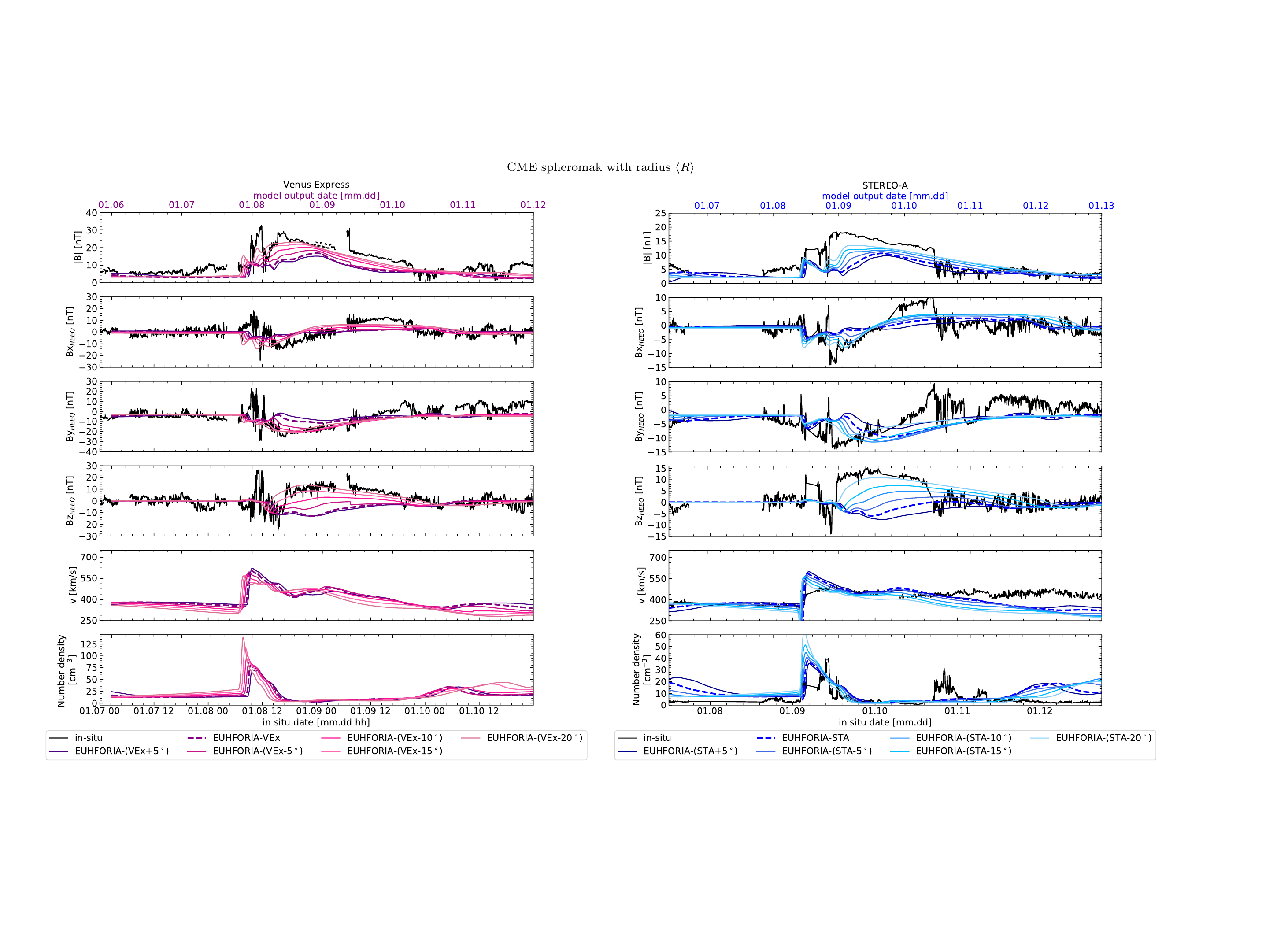}}
      \caption{Time series of the modelled plasma and magnetic field parameters at Venus Express and STEREO-A for a spheromak of size $\langle R \rangle$. The in situ observations are given for comparison purposes (black solid lines). The dashed coloured lines indicate the modelled time series at the exact location of the spacecraft, while the solid coloured lines are for the virtual spacecraft. All modelled curves were shifted by 1d4hr for Venus Express and 1d2hr for STEREO-A.
              }
         \label{timeseries_average}
   \end{sidewaysfigure*}

\begin{sidewaysfigure*}
   \resizebox{1\hsize}{!}
            {\includegraphics[trim=0 2.2cm 2cm 0, clip]{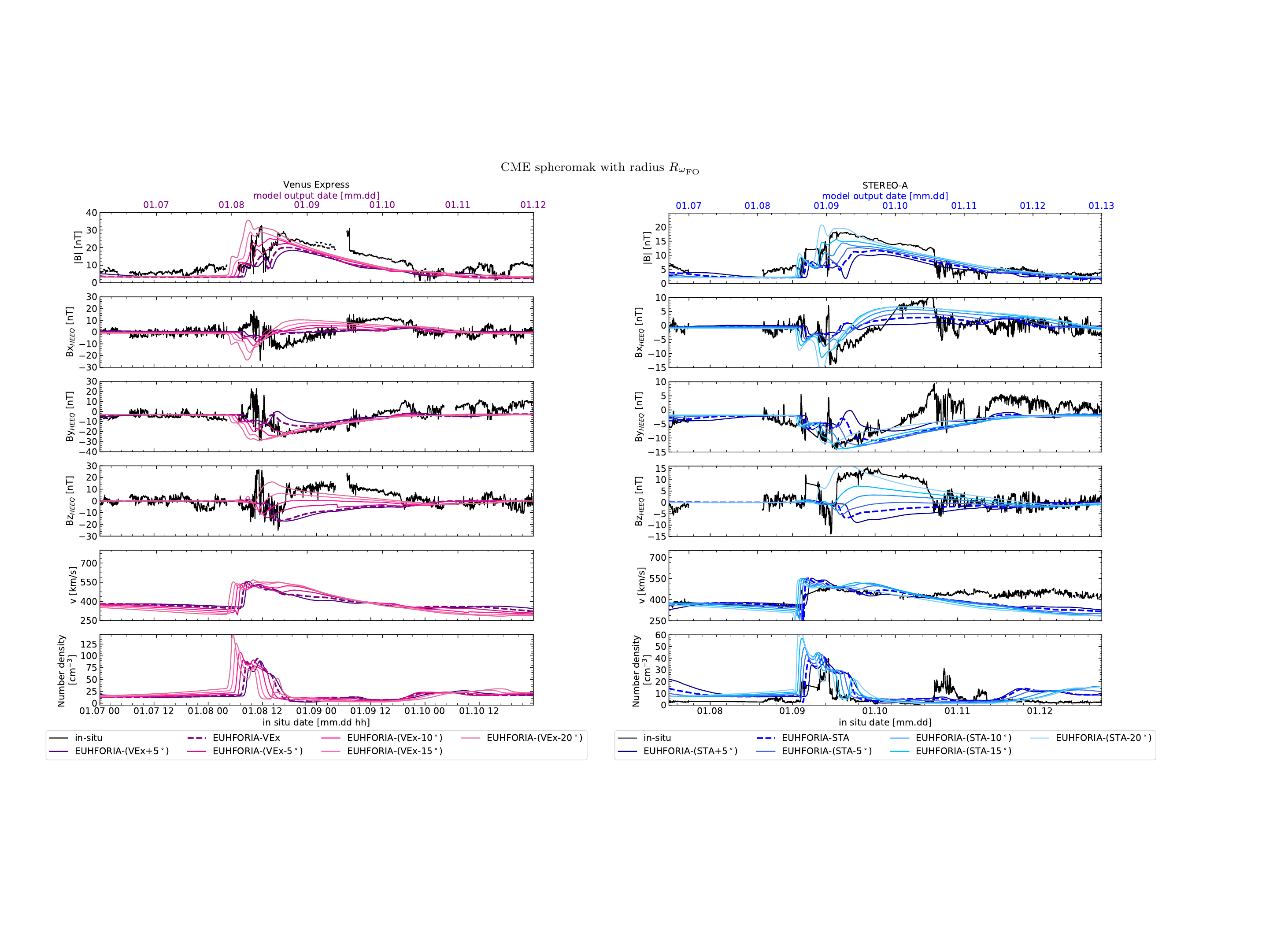}}
      \caption{Time series of the modelled plasma and magnetic field parameters at Venus Express and STEREO-A for a spheromak of size $R_{\mathrm{\omega_{FO}}}$. The in situ observations are given for comparison purposes (black solid lines). The dashed coloured lines indicate the modelled time series at the exact location of the spacecraft, while the solid coloured lines are for the virtual spacecraft. All modelled curves were shifted by 18hr for Venus Express and 19hr for STEREO-A.
              }
         \label{timeseries_large}
   \end{sidewaysfigure*}

Figures \ref{timeseries_small}, \ref{timeseries_average}, and \ref{timeseries_large} give the time series of the magnetic field and plasma properties registered in situ (black curve) and modelled respectively at Venus Express (dark purple dashed line marked as EUHFORIA-VEx in the legend) and STEREO-A (dark blue dashed line marked as EUHFORIA-STA) for all the spheromak CME runs. The solid lines show the results for virtual spacecraft placed at different locations. We first discuss the model outputs at the locations of STEREO-A and Venus Express. The model predicts the arrival of the CME earlier than the in situ observations. In the figures, to aid the comparison, the modelled results were shifted to match the in situ observed shock arrival at the two spacecraft. For each model run, the curves were shifted by the same amount as indicated in the figure legends. The dark purple and blue dashed lines show the model results at the locations of Venus Express and STEREO-A spacecraft, respectively.

The shock and flux rope arrival times for the EUHFORIA output for all three different spheromak sizes modelled ($R_{\mathrm{\omega_{EO}}} = 8.4R_\odot$, $R_{\mathrm{\omega_{FO}}} = 17.5R_\odot$, and $\langle R \rangle = 13.3R_\odot$) are given in Table \ref{table4}, together with the time difference between modelled and observed arrival times. The shock arrival is identified by the first increase in the magnetic field |B|, which coincides with the plasma speed and number density increases. For the flux rope in the simulation, the second increase in field |B| was accounted for as the flux rope start time. The time differences are defined as $\Delta t_{s} = t_{s model} -t_{s in situ}$ for the shock and $\Delta t_{f} = t_{f model} -t_{f in situ}$ for the flux rope, where $t_{s model}$ and $t_{f model}$ are the shock and the flux rope arrival times in the modelled time series, respectively, and $t_{s in situ}$ and $t_{f in situ}$ are the shock and flux rope arrival times in situ. The shock in the model output arrives at Venus Express within 5 -- 20 hours ahead of the observed in situ signature, while the flux rope leading edge arrives within 3 -- 15 hours earlier than the observed structure. For STEREO-A, the time difference interval is between 10 -- 25 hours for the shock and 6 -- 18 hours for the flux rope leading edge. These values are within the error bars set from other models \citep{mays_ensemble_2015}. The modelled structures are systematically faster than expected, with the best model result being the one from the largest spheromak modelled, which had initial radius of $R_{\mathrm{\omega_{FO}}}$. The larger the spheromak is, the shorter the identified sheath region becomes. The modelled sheath region is longer at STEREO-A compared to Venus Express. This happens regardless of the size of the modelled spheromak.

\begin{sidewaysfigure*}
   \centering
   \resizebox{0.75\hsize}{!}
            {\includegraphics[trim=0 100 250 100, clip]{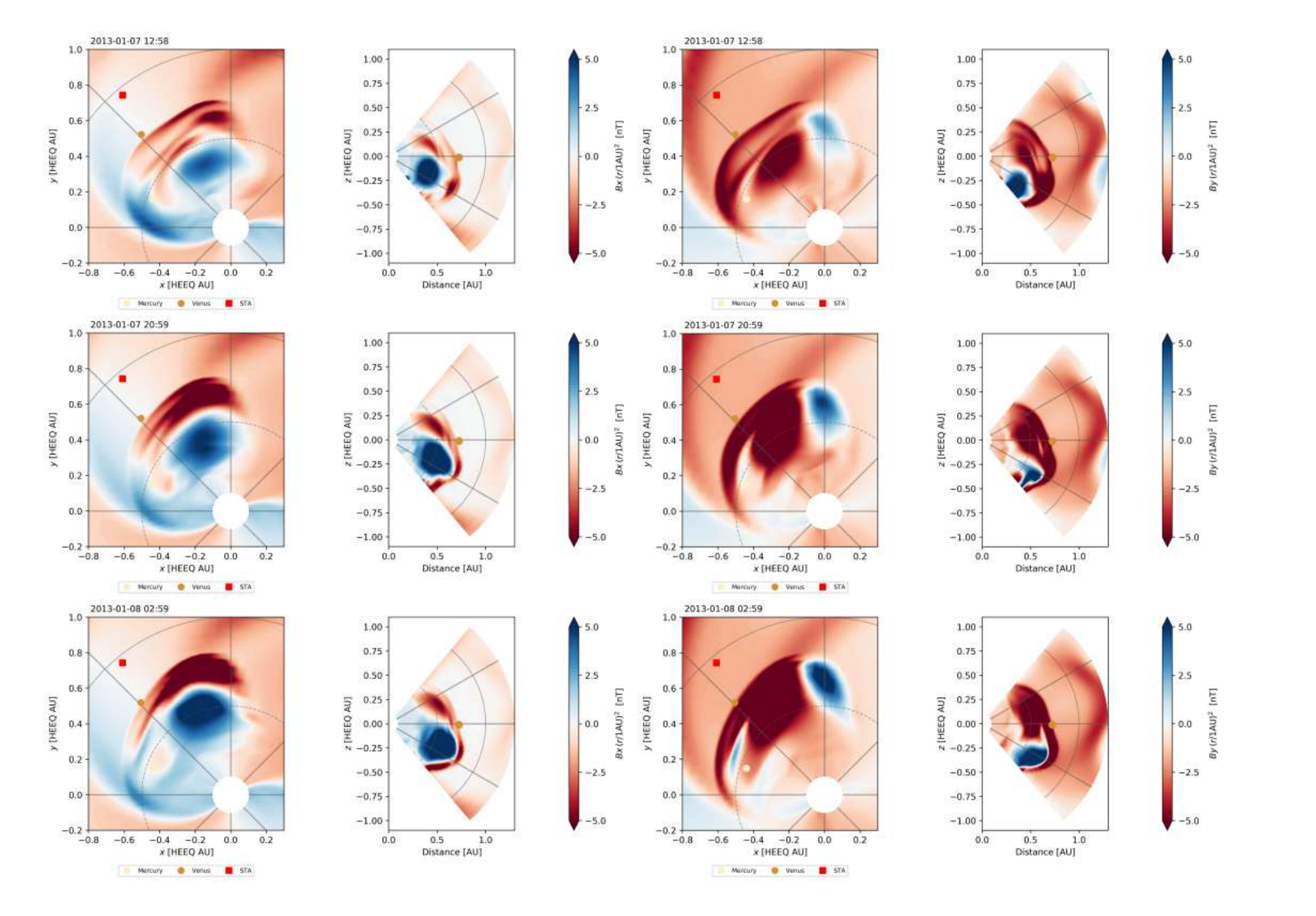}}
      \caption{Spatial profiles of the modelled $B_x$ and $B_y$ of the magnetic field of the CME upon arrival at Venus Express orbit. The order of placement is based on increasing radius of the spheromak CME from the top to bottom row ($R_{\mathrm{\omega_{EO}}}$, $\langle R \rangle$, and $R_{\mathrm{\omega_{FO}}}$).
              }
         \label{VEX_B_comp_eq_mer_minus90}
   \end{sidewaysfigure*}

We can see that the model output does not accurately capture the observed magnetic field. To further examine the flux rope and its manifestation at southern latitudes, we examined the output at virtual spacecraft placed at the same orbital radius as Venus Express and STEREO-A, but at different latitudinal positions. The virtual spacecraft were displaced by $-5^{\circ}$, $-10^{\circ}$, $-15^{\circ}$, and $-20^{\circ}$ in latitude from the actual position of the spacecraft. We also added one virtual spacecraft north of the real spacecraft at $+5^{\circ}$ in order to analyse the CME portion above. In Figures \ref{timeseries_small}, \ref{timeseries_average}, and \ref{timeseries_large}, the modelled time series at the virtual spacecraft are presented using solid coloured lines. It is clear from comparing EUHFORIA output time series to the in situ ones that the agreement improves the further south the virtual spacecraft is located. Comparing the solutions from the different spheromak CME sizes considered, it appears that the model output based on the largest of all the CMEs tested ($R_{\mathrm{\omega_{FO}}}$) shows better agreement with the in situ magnetic field signatures, especially for the virtual spacecraft located $-15^{\circ}$ and $-20^{\circ}$ from Venus Express and STEREO-A in latitude. In particular, the EUHFORIA output captures the $B_z$-component at the low latitude virtual spacecraft very well. The negative to positive rotation of the $B_x$- and $B_y$- components are also captured, as well as the magnitude and overall shape of the magnetic field magnitude profile. In addition, the two larger spheromak model the duration of the sheath region better; while for the smallest radius spheromak, the sheath is longer, resulting in a relatively large difference in the flux rope leading edge times. Similar improvements in the match between the model and in situ time series can also be obtained for virtual spacecraft positioned further east in longitude relative to the Sun--Venus Express--STEREO-A line. This can also be seen in the meridional and equatorial cuts of the B-components shown in Figures \ref{VEX_eq_mer_minus90}, \ref{STA_eq_mer_minus90}, and \ref{VEX_B_comp_eq_mer_minus90}. It is important to highlight, however, that the stronger fields in the time series produced by the larger in radius spheromak are due to the spacecraft intersecting a different part of the structure compared to the other model runs. The CME is injected in the modelling domain at 120.7$^{\circ}$ in longitude and –7.8$^{\circ}$ in latitude, and thus it does not propagate directly along the Sun - Venus Express - STEREO A line. The magnetic field is not evenly distributed among the volume of the spheromak, thus, it is expected that the line through which the spacecraft crosses at each simulation run intersects a slightly different part of the flux rope resulting in a different |B|. A larger sized spheromak would have a larger nose area of the CME compared to a smaller spheromak and thus there would be more chances that the spacecraft would travel through a strong |B| field.

It is worth recalling from remote-sensing observations that the CME appeared to propagate southward, away from the ecliptic plane. However, fitting of the in situ signatures from the near-ecliptic spacecraft to cylindrical flux rope models indicate a central flux rope encounter, that is to say the spacecraft crossed through or near the axis of the flux rope. More precisely, from observations, the fitted [$\theta$, $\phi$] axis directions relative to the x-y SpaceCraft EQuatorial (SCEQ) plane in degrees were [47, 165] at Venus Express and [36, 172] at STEREO-A \citep[see][for a detailed analysis]{good_self-similarity_2019}. Thus, the cylindrical flux rope is nearly aligned with the anti-sunward direction ($\phi$=180), but with quite a large out-of-ecliptic tilt (the $\theta$ values). Impact parameters are quite low (i.e. the spacecraft cut through the flux rope close to the axis), with values of 0.06 and 0.28 at Venus Express and STEREO-A, respectively. Bearing in mind that these fits only give the local axis direction, the results are consistent with either the flux rope having a high inclination and being globally tilted out of the ecliptic (therefore the spacecraft crossed closer to apex) or having a low inclination and lying more parallel to the ecliptic (therefore the spacecraft crossed through flux rope leg). Considering that the remote-sensing observations determine the initial CME parameters used for running EUHFORIA, we expect the modelled CME to follow a southern propagation. This was already indicated by the spatial profiles given in Figures \ref{VEX_eq_mer_minus90} and \ref{STA_eq_mer_minus90}. A southward propagation of the model CME is also consistent with the synthetic time series of the magnetic field and plasma parameters at the exact spacecraft locations.

\begin{table*}
\caption{Varying CME input parameters for the three spheromak runs.}
\centering
\begin{tabular}{c c c c c c c c c c c } 
\hline\hline\\
passage & & HEEQ & HEEQ & mass & & & & & &\\
 at 21.5$R_\odot$ & speed & longitude & latitude & density & temperature & radius & tilt & helicity & flux \\[1pt]
[UT] & [km/s] & [deg] & [deg] & [$kg/m^{3}$] & [K] & [$R_\odot$] & [deg] & sign & [Wb]\\[5pt]
\hline\hline\\[1pt]
6 January 2013 12:49:33 & 571.0 & 120.7 & -7.8 & 1e-18 & 0.8e6 & 8.4 & -40.0 & +1 & 80.0e12 & \\[3pt]
6 January 2013 12:49:33 & 571.0 & 120.7 & -7.8 & 1e-18 & 0.8e6 & 17.5 & -40.0 & +1 & 80.0e12 & \\[3pt]
6 January 2013 12:49:33 & 571.0 & 120.7 & -7.8 & 1e-18 & 0.8e6 & 13.3 & -40.0 & +1 & 80.0e12 & \\[3pt]
\hline  
\label{table3}
\end{tabular}
\end{table*}

\section{Conclusions}

In this study, we analysed a multi-spacecraft CME encounter in order to investigate whether the EUHFORIA model can reconstruct the evolution of the flux rope observed by two radially aligned spacecraft. The event, an extended filament eruption that occurred on 6 January 2013, created clear in situ signatures at Venus Express and STEREO-A, while it only produced a mild disturbance at MESSENGER. The location of the eruption did not allow for information about the magnetic flux content of the CME to be extracted. However, EUV and white-light images allowed for the determination of the eruption time and the CME helicity sign, as well as the extraction of kinetic and geometric parameters of the large-scale structure using the GCS model. Although two more eruptions occurred close to the event under study, we deduced from the analysis of an observational and from a modelling assessment that neither of these eruptions  significantly interfered with the primary event in this study. The CME was modelled using three different sets of parameters. The three simulation runs were done for different radii for the spheromak ($R_{\mathrm{\omega_{EO}}}$, $\langle R \rangle$, and $R_{\mathrm{\omega_{FO}}}$). Based on the orientation of the PIL and spheromak, we justified that it is necessary to rotate the spheromak by 90 degrees anticlockwise from the GCS-derived tilt, and we submitted the runs using this modified tilt. The main conclusions of the observational and modelling analysis are as follows.

   \begin{enumerate}
      \item Magnetic field and plasma measurements by Venus Express and STEREO-A suggested that the observed CME did not significantly evolve between the two spacecraft measurements.
      
      \item Remote-sensing observations indicated a direction of propagation pointing southward.
      
      \item Matching the GCS fitted structure to that of the spheromak used in EUHFORIA is not unique and can be difficult, introducing different possibilities reflecting the fact that the magnetic field structure is not directly manifested in the white-light emission and thus poorly constrained.
      
      \item The radius of the modelled spheromak had an impact on the modelled magnetic field profiles and their amplitude, the arrival times, and the distance between the shock and the flux rope arrival.
      
      \item Similarly to what was indicated by the white-light images, the model showed that the direction of propagation of the CME was mainly southward. The modelled time series at the real spacecraft locations did not fully agree with the in situ measurements, while virtual spacecraft placed at the same radial distances but lower in latitude showed better agreement to the in situ observations. The same holds true for a longitudinal displacement further east relative to the Sun--Venus Express--STEREO-A line. This suggests that a better agreement between observations and modelled result is found in these areas.
      
      \item One possible explanation is that the observed CME underwent deflection from its initial course in the inner heliosphere \citep{zuccarello_role_2012}. The possibility of pancaking of the CME, which would have resulted in flattening and stretching, and which are not modelled by the simulation, may also be a  contributing factor. In this case, despite a more southward propagation of the CME, the flux rope could still have been stretched to higher latitudes, resulting in a more nose encounter of the CME with the two spacecraft. This would be in accordance with the in situ signatures that indicate an encounter closer to the nose. This kind of effect would not have necessarily been captured by EUHFORIA, at least not to a full extent especially if the front flattening took place before 0.1 AU.
      
      \item The predicted arrival times are well ahead of the ones extracted from the in situ signatures but within errors previously established from other studies and possibly a result of both the propagation and the expansion speed being included in the GCS fitting approach.
   \end{enumerate}

   The aim of this work was to assess the degree to which the EUHFORIA--spheromak model is able to capture the evolution of CMEs in the context of multi-spacecraft encounters. Although there were some unexpected inaccuracies in the output, we conclude that the model predicted the B-field and most importantly the $B_z$-component at the nose of the CME well, which appeared to have been encountered in situ by Venus Express and STEREO-A. This is of crucial importance when studying Earth-directed CMEs. Of course here we analysed one event, so generalising that EUHFORIA--spheromak will always be accurate in the $B_z$-field predictions would be misleading. Further investigation is required for such conclusions.

\begin{acknowledgements}
      This work was completed under the Project TRAMSEP (Transport Mechanisms of Strong Solar Energetic Particles in Complex Background Solar Wind Conditions), personal funding of E. Asvestari (Academy of Finland Grant 322455). EUHFORIA is developed as a joint effort between the University of Helsinki and KU Leuven. The validation of solar wind and CME modelling with EUHFORIA is being performed within the BRAIN-beproject CCSOM (Constraining CME and Shocks by Observations and Modelling throughout the inner heliosphere; http://www.sidc.be/ccsom/). The results presented here have been achieved under the framework of the Finnish Centre of Excellence in Research of Sustainable Space (Academy of Finland Grant312390), which we gratefully acknowledge. J. P. , E. K. and S. P. acknowledge funding from the European Union Horizon 2020 research and innovation program under grant agreement 870405 (EUHFORIA 2.0). E. P. acknowledges the NASA Living With a Star Jack Eddy Postdoctoral Fellowship Program, administered by UCAR's Cooperative Programs for the Advancement of Earth System Science (CPAESS) under award no. NNX16AK22G. These results were also obtained in the framework of the projects C14/19/089  (C1 project Internal Funds KU Leuven), G.0D07.19N  (FWO-Vlaanderen), and C~90347 (ESA Prodex). This work benefited from open access to GONG magnetograms developed with the ADAPT model, and the EUV filtergrams and white-light images obtained by instruments on board SDO provided by JSOC, and STEREO-A and STEREO-B. E. The Kanzelh\"ohe H$\alpha$ data were provided by the Kanzelh\"ohe Solar Observatory, University of Graz, Austria. E. Asvestari is ever grateful to Dr. Tobias Rindlisbacher for his moral support through the process of writing this paper. We would like to thank the referee, Dr. Ward  Manchester, for thoroughly reviewing our manuscript and for the useful feedback.
\end{acknowledgements}

\begin{table*}
\caption{EUHFORIA shock and flux rope arrival times at Venus Express and STEREO-A (model result at the ecliptic - 0 - latitude plane) and the time difference between modelled and in situ arrival times for the shock ($\Delta t_{s} = t_{s model} -t_{s in situ}$) and the flux rope ($\Delta t_{f} = t_{f model} -t_{f in situ}$), where $t_{s model}$ and $t_{f model}$ are the shock and the flux rope arrival times in the modelled time series, respectively, and $t_{s in situ}$ and $t_{f in situ}$ are the in situ shock and flux rope arrival times accordingly.}
\centering
\begin{tabular}{c c c c c c} 
\hline\hline\\
Spheromak size & spacecraft & shock arrival [UT] & Flux rope arrival [UT] & $\Delta t_{s}$ & $\Delta t_{f}$ \\[2pt]
\hline\hline\\[1pt]  
\multirow{2}{*}{$R_{\mathrm{\omega_{EO}}}$} & Venus Express & 7 January 2013 13:30 & 8 January 2013 00:00 & 19h52min & 15h24min\\[3pt]
& STEREO-A & 8 January 2013 00:30 & 8 January 2013 16:00 & 25h55min & 18h39min\\[3pt]
\hline\\[1pt]
\multirow{2}{*}{$\langle R \rangle$} & Venus Express & 7 January 2013 22:30 & 8 January 2013 07:00  & 10h52min & 8h24min\\[3pt]
& STEREO-A & 8 January 2013 12:00 & 9 January 2013 00:00 & 14h25min & 10h39min\\[3pt]
\hline\\[1pt]
\multirow{2}{*}{$R_{\mathrm{\omega_{FO}}}$} & Venus Express & 8 January 2013  03:30 & 8 January 2013 11:45 & 5h52min & 3h39min\\[3pt]
& STEREO-A & 8 January 2013 15:30 & 9 January 2013 04:30 & 10h55min & 6h9min\\[3pt]
\hline  
\label{table4}
\end{tabular}
\end{table*}

   \bibliographystyle{aa}
   \bibliography{Multispacecraft_encounter.bib} 

\begin{appendix}

\section{Excluding an interacting CMEs scenario}\label{cone_EUHFORIA}

\begin{table*}
\caption{Varying CME input parameters for the cone EUHFORIA run discussed in the Appendix \ref{cone_EUHFORIA}.}
\centering
\begin{tabular}{c c c c c c c c} 
\hline\hline\\
CME & passage at 21.5$R_\odot$ & speed & longitude & latitude & half width & mass density & temperature\\
& [UT] & [km/s] & [deg HEEQ] & [deg HEEQ] & [deg] & [$kg/m^{3}$] & [K]\\[3pt]
\hline\hline\\  
CME 1 & 6 January 2013 15:06:13 & 399.8 & 155.6 & -10.2 & 11.2 & 1e-18 & 0.8e6 \\[3pt]
CME 2 & 6 January 2013 12:49:33 & 571.0 & 120.7 & -7.8 & 31.6 & 1e-18 & 0.8e6 \\[3pt]
\hline\hline\\[1pt]
\label{table_append}
\end{tabular}
\end{table*}

As mentioned in Section \ref{remotesensing}, a neighbouring filament erupted approximately 25 hours prior to the main event under study. In order to exclude the possibility of the two events interacting in interplanetary space, we first modelled both eruptions using the cone CME model implemented in EUHFORIA. To get the CME morphology and the kinematics of this earlier erupting neighbouring filament, we applied the GCS model in the same manner as described in Section \ref{mor_kin}. The input parameters for the cone CME runs are given in the first two rows of Table \ref{table_append}, where by CME 1 is meant to be the earlier erupting filament and CME 2 is the main event studied.

\begin{figure}
   \centering
   \includegraphics[width=0.95\hsize]{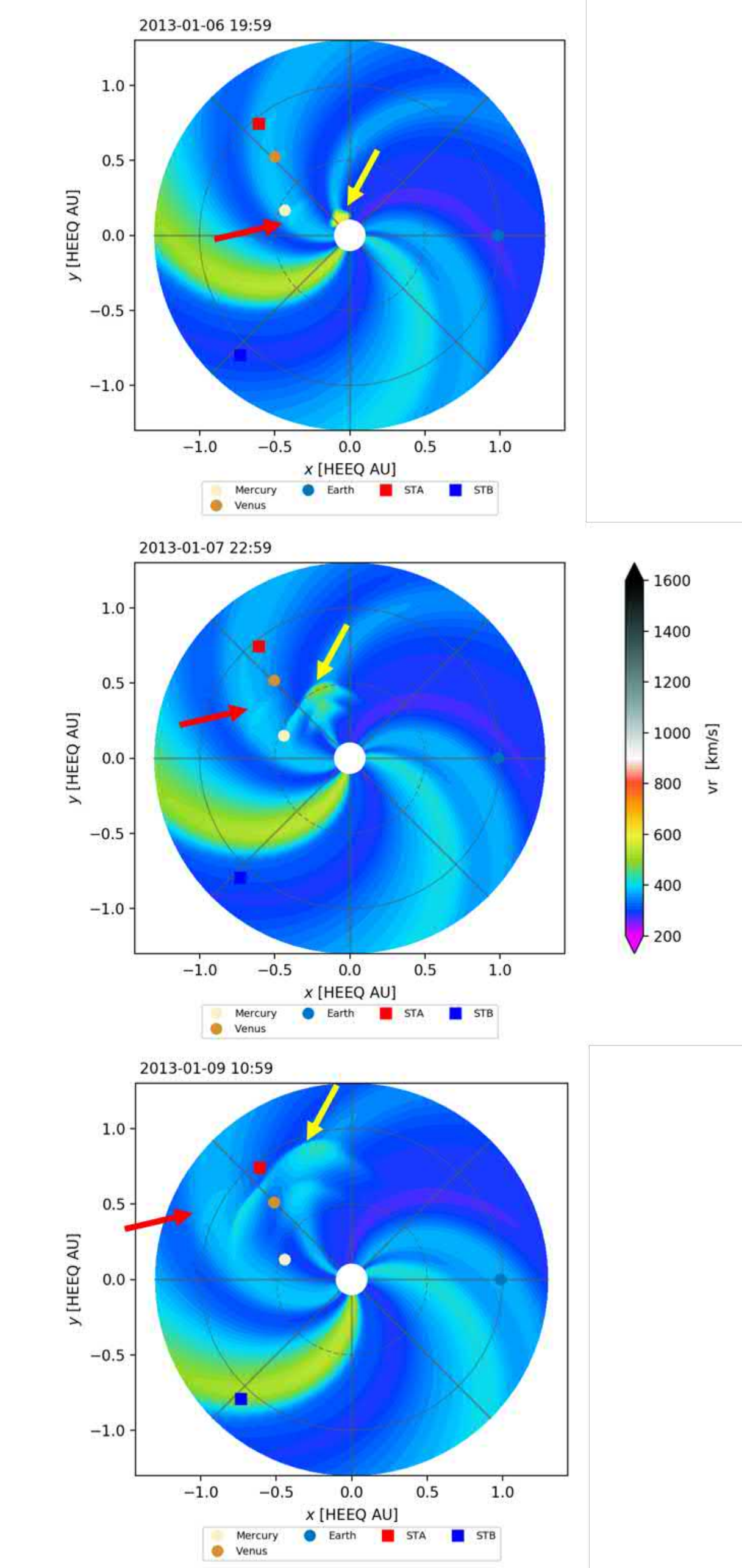}
      \caption{Equatorial cuts of the EUHFORIA output modelling the two filament eruptions. The red arrow in each panel indicates the modelled CME that corresponds to the filament that erupted first and the yellow arrow shows the second CME that would be generated by the filament eruption that is the primary event in this study.
              }
         \label{cone_model}
   \end{figure}

Figure \ref{cone_model} shows equatorial cuts of the spatial distribution of the radial velocity from the model output for the double cone CME case. Each panel gives the model result at a different timestamp ordered from an earlier to a later one from top to bottom. The top image is a snapshot of the model 7 hours after both CMEs were inserted. At that time, CME 1 had already reached Mercury (planet shown as a yellow disc). It is a faint structure indicated by a red arrow. The second eruption, CME 2, was still relatively close to the Sun (indicated by a yellow arrow). It is already clear that the first filament generated a very weak CME in terms of its speed, with a very narrow longitudinal extent on the equatorial plane and a direction of propagation that would only result in a leg encounter with Venus Express and STEREO-A (planets shown as an orange disc and red square, respectively), as can be seen from the later snapshots. The two CMEs do not appear to catch up within the modelling domain. This is in accordance with observations since, as discussed in Section \ref{remotesensing}, there are no signs of interacting structures in the in situ signatures. We therefore conclude that is sufficient to consider the second filament eruption as an isolated event.

\end{appendix}

\end{document}